\documentclass[%
 reprint,
superscriptaddress,
 amsmath,amssymb,
 aps,
 prl,
]{revtex4-2}

\usepackage{graphicx}
\usepackage{dcolumn}
\usepackage{bm}
\usepackage[mathlines]{lineno}
\usepackage{braket}
\usepackage[dvipsnames]{xcolor}
\usepackage{amsmath,amssymb,amsfonts} 
\usepackage{mathrsfs} 
\usepackage{appendix}
\usepackage{mathtools}
\usepackage{makecell}
\usepackage{float}
\usepackage{bbm}
\usepackage{nicefrac}
\usepackage{wasysym}
\usepackage[normalem]{ulem}

\usepackage{hyperref}
\hypersetup{
	colorlinks=true,
	linkcolor=blue,
	filecolor=blue,
	urlcolor=blue,
	citecolor=blue,
}

\def\beq{\begin{equation}}
\def\eeq{\end{equation}}
\def\bea{\begin{eqnarray}}
\def\eea{\end{eqnarray}}
\def\ket#1{\vert#1\rangle}
\def\bra#1{\langle#1\vert}

\def\rr{{\bm r}}

\def\kk{{\bm k}}

\begin{document}

\title{Orbital Altermagnetism in Two Dimensions}

\author{Mingxiang Pan}
\affiliation{School of Physics, Peking University, Beijing 100871, China}

\author{Feng Liu}
\email[Contact author: ]{ftiger.liu@utah.edu}
\affiliation{Department of Materials Science and Engineering, University of Utah, Salt Lake City, Utah 84112, USA}

\author{Huaqing Huang}
\email[Contact author: ]{huaqing.huang@pku.edu.cn}
\affiliation{School of Physics, Peking University, Beijing 100871, China}
\affiliation{Collaborative Innovation Center of Quantum Matter, Beijing 100871, China}
\affiliation{Center for High Energy Physics, Peking University, Beijing 100871, China}

\date{\today}

\begin{abstract}
We introduce the concept of \emph{orbital altermagnetism}, a symmetry-protected magnetic order of pure orbital degrees of freedom. It is characterized with ordered anti-parallel orbital magnetic moments in real space but momentum-dependent orbital band splittings, analogous to spin altermagnetism. Using a minimal tight-binding model with complex hoppings in a square-kagome lattice, we show that such order inherently arises from staggered loop currents, producing a $d$-wave-like orbital-momentum locking. First-principles calculations show that orbital altermagnetism emerges independent of spin ordering in in-plane ferromagnets of CuBr$_2$ and VS$_2$, so that it can be unambiguously identified experimentally. On the other hand, it may also coexist with spin altermagnetism, such as in monolayer MoO and CrO. The orbital altermagnetism offers an alternative platform for symmetry-driven magnetotransport and orbital-based spintronics, as exemplified by large nonlinear current-induced orbital magnetization.
\end{abstract}
\maketitle

\textit{Introduction.}---Altermagnetism has recently emerged as a distinct class of magnetic order, fundamentally different from both conventional collinear ferromagnetism and antiferromagnetism~\cite{PhysRevX.12.031042, PhysRevX.12.040501,sciadv.aaz8809,JunweiLiu2021Natcomm, Song2025NatureRevMater, adfm.202409327}. Characterized by symmetry-protected, momentum-dependent alternating spin splitting in the band structure and collinear compensated magnetic moments in real space, altermagnets enable novel spin transport phenomena, such as giant and tunnelling magnetoresistance \cite{PhysRevX.12.011028}, nonrelativistic spin current generation \cite{Naka2019NC, PhysRevLett.126.127701, Bose2022NatEle,  PhysRevLett.130.216701}, and nonlinear transports \cite{PhysRevLett.133.106701, PhysRevB.111.125420}.
These properties make them promising platforms for spintronics, offering high-speed and low-dissipation spin control without a net magnetization~\cite{Naka2019spincurrent, Tanaka_2025, PhysRevX.14.011019, Kremps2024Nature, Takagi2025Spontaneous, Reichlova2024Observation, PhysRevLett.132.036702, PhysRevLett.134.106802, PhysRevX.15.021083, PhysRevLett.131.256703, pnas.2108924118, v3fg-6smc}.

While magnetic order is usually pertained to spin degrees of freedom, orbital contributions to magnetism can be equally important. Ferromagnetic order formed by orbital magnetic moments has been reported in mori\'e Chern insulators \cite{science.abd3190, science.aay5533, science.abh2889, Nature2020orbitalChern, Sharpe2021Nanolett, acs.nanolett.1c03699, xie2025unconventionalorbitalmagnetism} and rhombohedral graphene superlattices \cite{Julong2023Nature_rhomb, science.adk9749, Luxiaobo2025Natmat_rhomb, PhysRevB.107.L121405, PhysRevB.109.L060409, PhysRevB.111.165102}. Weak orbital ferromagnetism can even survive in spin antiferromagnets \cite{PhysRevLett.76.4963, PhysRevLett.87.116801, JPSJ.86.063703, PhysRevLett.134.196703}. Another prominent example is the interaction-driven loop-current phase proposed in cuprates \cite{nersesyan1989low, Varma2010Nature, PhysRevB.55.14554, PhysRevLett.83.3538, PhysRevB.73.155113, PhysRevB.80.214501, Varma_2014, PhysRevB.92.195140, PhysRevLett.112.017004}, where spontaneous circulating currents within a unit cell generate finite orbital moments. In a mean-field description, such states correspond to complex inter-orbital hopping amplitudes, breaking time-reversal symmetry $\mathcal{T}$ and producing orbital antiferromagnetism independent of spin ordering. Also, interaction driven ordering has been proposed to interestingly promote spin altermagnetism \cite{PhysRevLett.132.236701}. These developments naturally raise a fundamental question: can local orbital magnetic moments \emph{alone} form an altermagnetic state, in analogous to spin altermagnetism?

In this Letter, we introduce the concept of \emph{orbital altermagnetism}---an altermagnetic order originating from staggered local orbital magnetic moments, namely, a magnetic ordered phase of pure orbital degrees of freedom \cite{yu_altermagnetism_2025, PhysRevResearch.7.023152, PhysRevLett.134.146001}. Using a minimal tight-binding model with spontaneously ordered loop currents, we demonstrate the emergence of a characteristic real-space pattern of anti-parallel orbital magnetic moments and a distinct feature of \textit{d}-wave-like orbital-momentum locking in momentum space.  Importantly, we show that orbital altermagnetism can exist independent of spin configurations in realistic materials. First-principles calculations reveal a definite manifestation of orbital altermagnetism in in-plane ferromagnets CuBr$_2$ and VS$_2$ with staggered out-of-plane orbital moments distinct from the uniform in-plane spin distributions. Furthermore, we find that conventional spin altermagnets can be accompanied with a hidden orbital altermagnetism, as exemplified by monolayer MoO and CrO, and we uncover a giant nonlinear current-induced orbital magnetization that far exceeds the spin contribution, which may serve as an experimental signature of the orbital altermagnetism.

\begin{figure*}
\centering
\includegraphics[width=0.8\linewidth]{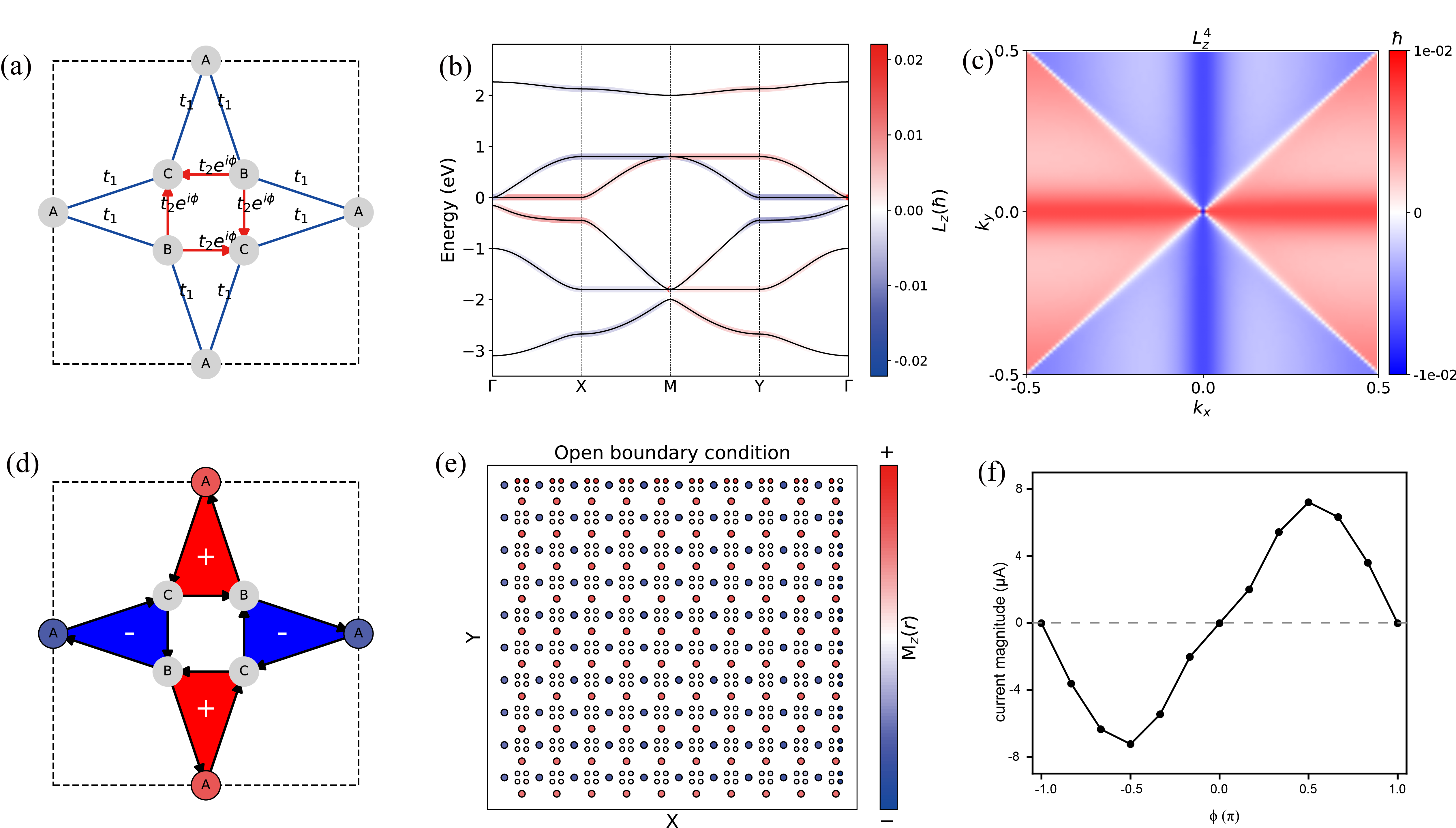}
\caption{\label{fig:lat}(a) Unit cell (dashed box) of the tight-binding model described by Eq.~(\ref{eq:ham}) includes three types of sites labeled A, B, and C in a square-kagome lattice, respectively. The nearest-neighbor hopping $t_2e^{i\phi}$ has a complex phase (red arrow), and next-nearest-neighbor hopping $t_1$ (blue) is real. Arrows indicate the direction of positive phase hopping. (b) Band structure of the model. (c) $\kk$-space distribution of the orbital angular momentum $L_z$ for the fourth band, with $k_x$ and $k_y$ in reduced coordinates. (d) Loop-current pattern in the model. Arrows indicate the calculated inter-site currents ($\sim5.4\, \mu$A), with red (+) and blue (-) regions denoting positive and negative $z$-components of the orbital magnetic moment $M_z$. (e) Real-space distribution of orbital magnetic moments in a $10\times 10$ lattice under open boundary conditions at $\mu=0$ eV. Red (blue) indicates a positive (negative) $M_z(\rr)$. (f) Inter-site current as a function of phase $\phi$. Model paramters: $\epsilon_A=-1.0$, $\epsilon_{B/C}=0$, $t_1=-0.6$, $t_2=-1.0$, and $\phi=\pi/3$. }
\end{figure*}

\textit{Minimal model for 2D orbital altermagnetism.}---We illustrate orbital altermagnetism using a minimal tight-binding model on a two-dimensional square-kagome lattice \cite{PhysRevB.65.014417, PhysRevLett.93.107206, PhysRevX.4.041037} with complex hoppings [Fig.~\ref{fig:lat}(a)]. Each unit cell contains three nonequivalent sites, labeled A, B, and C. The A sites sit at edge centers, $(\tfrac{1}{2},0)$ and $(0,\tfrac{1}{2})$, while B and C occupy positions of $(\tfrac12 \pm \tau,\tfrac12 \pm \tau)$ and $(\tfrac12 \pm \tau,\tfrac12 \mp \tau)$, respectively, with $\tau = 1/8$. The Hamiltonian reads
\begin{equation}
\label{eq:ham}
\begin{split}
H =& \sum_{\alpha,i} \epsilon_{\alpha} c_{\alpha i}^\dagger c_{\alpha i}+\sum_{\langle\langle i,j\rangle\rangle} t_1
(c_{Ai}^\dagger c_{Bj} + c_{Ai}^\dagger c_{Cj}) \\
& + \sum_{\langle i,j\rangle} t_2 e^{i\phi} c_{Ci}^\dagger c_{Bj} + \text{H.c.},
\end{split}
\end{equation}
where $c_{\alpha i}^\dagger$ ($c_{\alpha i}$) creates (annihilates) an electron on site $i$ of type $\alpha = A,B,C$. The first term describes next-nearest-neighbor hopping with amplitude $t_1$, and the second term corresponds to nearest-neighbor hopping of amplitude $t_2$ with an orbital-dependent phase $e^{i\phi}$. The phase $\phi$ encodes the microscopic loop current pattern responsible for orbital altermagnetic order. The magnetic point group of this model, with the orbital moment acting like spins, is $4^\prime/mm^\prime m$, as verified from symmetry operations acting on the lattice with directed bonds.

We first compute the electronic structure and the orbital angular momentum $L_z$, which is directly proportional to the orbital magnetic moment $M_z=-\mu_B g_L L_z$ ($g_L=1$) within the modern theory of orbital magnetization~\cite{PhysRevLett.95.137205, PhysRevB.74.024408, cphc.200400641, PhysRevLett.99.197202,  Resta_2010, OM_in_solids, RevModPhys.82.1959,PhysRevB.107.115102}.
As shown in Fig.~\ref{fig:lat}(b), the bands exhibit opposite $L_z$ polarization along $k_x$ and $k_y$. The momentum-space distribution of $L_z$ [Fig.~\ref{fig:lat}(c)] displays a characteristic $L_z$-momentum locking, analogous to the spin texture of $d$-wave spin altermagnets. The corresponding real-space orbital magnetization, computed from its Wannier-based expression~\cite{PhysRevLett.110.087202, PhysRevB.95.121114, PhysRevB.106.075136, PhysRevB.103.195309}, is shown in Fig.~\ref{fig:lat}(e) for an open-boundary system. The real-space orbital moments, which are decomposed to lattice sites, are predominantly localized on A sites and opposite orientation in a N\'{e}el-antiferromagnetic-like pattern across the lattice, yielding a fully compensated configuration. Therefore, in analogy to spin altermagnetism, we refer to this state as \emph{orbital altermagnetism}---characterized by a compensated real-space arrangement and alternating momentum-space orbital splitting. The observed $d$-wave pattern is fully compatible with the magnetic space group,  albeit with orbital-momentum locking in place of spin-momentum locking or spin-orbit coupling (SOC).

\begin{table*}[t!]
\centering
\caption{ {Two-dimensional magnetic point groups compatible with orbital altermagnetism.}}
\label{tab:mpg}
\begin{tabular}{c c c}
\hline\hline
 {Leading wave} &  {Allowed magnetic point groups (MPGs)}  \\ \\
\hline
 {\textit{p}-wave} &  {$\mathrm{1.1',\ 2.1,\ 2.1',\ m.1,\ m.1'}$} \\
 {\textit{d}-wave} &  {$\mathrm{2/m.1,\ 222.1,\ mm2.1,}$}\newline
          {$\mathrm{ mmm.1,\ 4',\ -4',\ 4'/m,\ 4'22',\ 4'm'm,\ -4'2'm,\ -4'2m',\ 4'/mm'm}$} \\
 {\textit{g}-wave} &  {$\mathrm{422.1,\ 4mm.1,\ -42m.1,\ 4/mmm}$} \\
 {\textit{f}-wave} &  {$\mathrm{3.1',\ 32.1,\ 32.1',\ 3m.1,\ 3m.1',\ -6.1',\ -6',\ -6m2.1,\ -6m2.1',\ -6'm'2,\ -6'm2'}$} \\
 {\textit{i}-wave} &  {$\mathrm{-3m.1,\ 622.1,\ 6mm.1,\ 6/mmm.1}$} \\
\hline\hline
\end{tabular}
\end{table*}

The underlying physics can be understood in terms of local loop currents. Figure~\ref{fig:lat}(d) shows the bond-resolved current distribution, where black arrows indicate current flow. On each ABC triangle, a current of magnitude $\sim5.4\; \mu$A circulates in the A-C-B direction, generating an out-of-plane orbital magnetic moment at the center of triangle. Neighboring triangles host counter-circulating loops, and hence antiparallel moments to produce an antiferromagnetic arrangement of orbital magnetic moments. Triangles with opposite orbital magnetization are related by $C_{4z}\mathcal{T}$ but not by $\mathcal{PT}$ or $\tau\mathcal{T}$, consistent with general symmetry requirements for altermagnetism \cite{PhysRevX.12.031042,PhysRevB.110.054406, Cheong2025Alterm_classification}. Based on the loop-current picture, we evaluate the magnetic multipole and find a nonzero magnetic octupole in the minimal model, which serves as a secondary order parameter in the Landau theory of $d$-wave altermagnetism \cite{PhysRevLett.132.176702} (see Supplementary Material (SM) for details~\footnote{\label{fn}See Supplemental Material at http://link.aps.org/supplemental/xxx, for more details about the expressions of orbital magnetization in reciprocal and real spaces, magnetic multipoles, the formula for nonlinear current-induced magnetization, detailed symmetry analysis of orbital altermagnetism, the first-principles calculation method, and more numerical result of material candidates, which includes Refs.~\cite{dww3-vm74,4kmy-59l9, PhysRevB.106.104414,PhysRevResearch.5.043052,PhysRevB.109.094438, 10.21468/SciPostPhys.18.3.109, liu_symmetry_2026, doi:10.7566/JPSJ.93.072001,PhysRevB.98.165110, doi:10.7566/JPSJ.87.033709, PhysRevB.104.054412, PhysRevB.107.195118, PhysRevLett.77.3865, KRESSE199615, 10.1063/5.0147450, CALDERON2015233, MOSTOFI2008685, PhysRevX.14.031037, cam5007}.}).
\nocite{dww3-vm74,4kmy-59l9, PhysRevB.106.104414, PhysRevResearch.5.043052,PhysRevB.109.094438, 10.21468/SciPostPhys.18.3.109, liu_symmetry_2026, doi:10.7566/JPSJ.93.072001,PhysRevB.98.165110, doi:10.7566/JPSJ.87.033709, PhysRevB.104.054412, PhysRevB.107.195118, PhysRevLett.77.3865, KRESSE199615, 10.1063/5.0147450, CALDERON2015233, MOSTOFI2008685, PhysRevX.14.031037, cam5007}
Furthermore, we find that the current magnitude $|I|$ and direction are strongly phase-dependent: as shown in Fig.~\ref{fig:lat}(f), $|I|$ varies as $\sin\phi$ due to time-reversal constraints, reversing direction upon $\phi \to -\phi$, with a maximum magnitude at $\phi=\pm\pi/2$ and vanishing at $\phi=0\;\mathrm{or}\;\pm\pi$, where the hopping becomes real. Finally, while loop currents provide a minimal realization with an intuitive physical picture in the present model, orbital altermagnetism is, in general, uniquely characterized by a symmetry-enforced alternating pattern of orbital magnetization and does not necessarily require a loop-current origin. Thus, the existence of two-dimensional orbital altermagnetism is ultimately dictated by symmetry rather than by a specific microscopic mechanism, and its compatible magnetic point groups are summarized in Table~\ref{tab:mpg} (the detailed screening procedure is given in Sec.~V of the SM~\footnotemark[\value{footnote}]).

Before proceeding to real orbital altermagnetic materials, we make a few remarks on the general physics of orbital altermagnetism. (i) Orbital antiferromagnetism has been studied in interaction-driven loop-current states in correlated systems such as underdoped cuprates \cite{PhysRevLett.89.247003, PhysRevB.63.094503, PhysRevB.64.012502, S0217979201007002, PhysRevB.66.012505, PhysRevB.72.132501}, SrRuO$_3$ \cite{PhysRevB.69.094407}, CeB$_6$ \cite{JPSJ.54.3909}, and URu$_{2-x}$Fe$_x$Si$_2$ \cite{PhysRevLett.117.227601}, but its realization as an \emph{altermagnetic} order remains unexplored. (ii) Our proposed orbital altermagnetism is distinct from orbital order~\cite{PhysRevLett.132.236701} and orbital-spin locking~\cite{bzzy-ngcs}, where the staggered orbital ordering is leveraged to produce spin altermagnetism via breaking sublattice equivalence. (iii)
Since orbital magnetization is often induced by effective complex hoppings via SOC or correlation effects, the symmetry must be described by magnetic groups rather than spin groups. Since symmetry required for orbital and spin altermagnetism is the same, they may coexist when orbital moments are aligned with spins or collectively shifted by a fraction of the period of the lattice. On the other hand, however, orbital altermagnetism may occur independent of spin ordering including spin altermagnetism when spins are absent or oriented differently, such as in ferromagnetic states.

\begin{figure}
\centering
\includegraphics[width=1.0\linewidth]{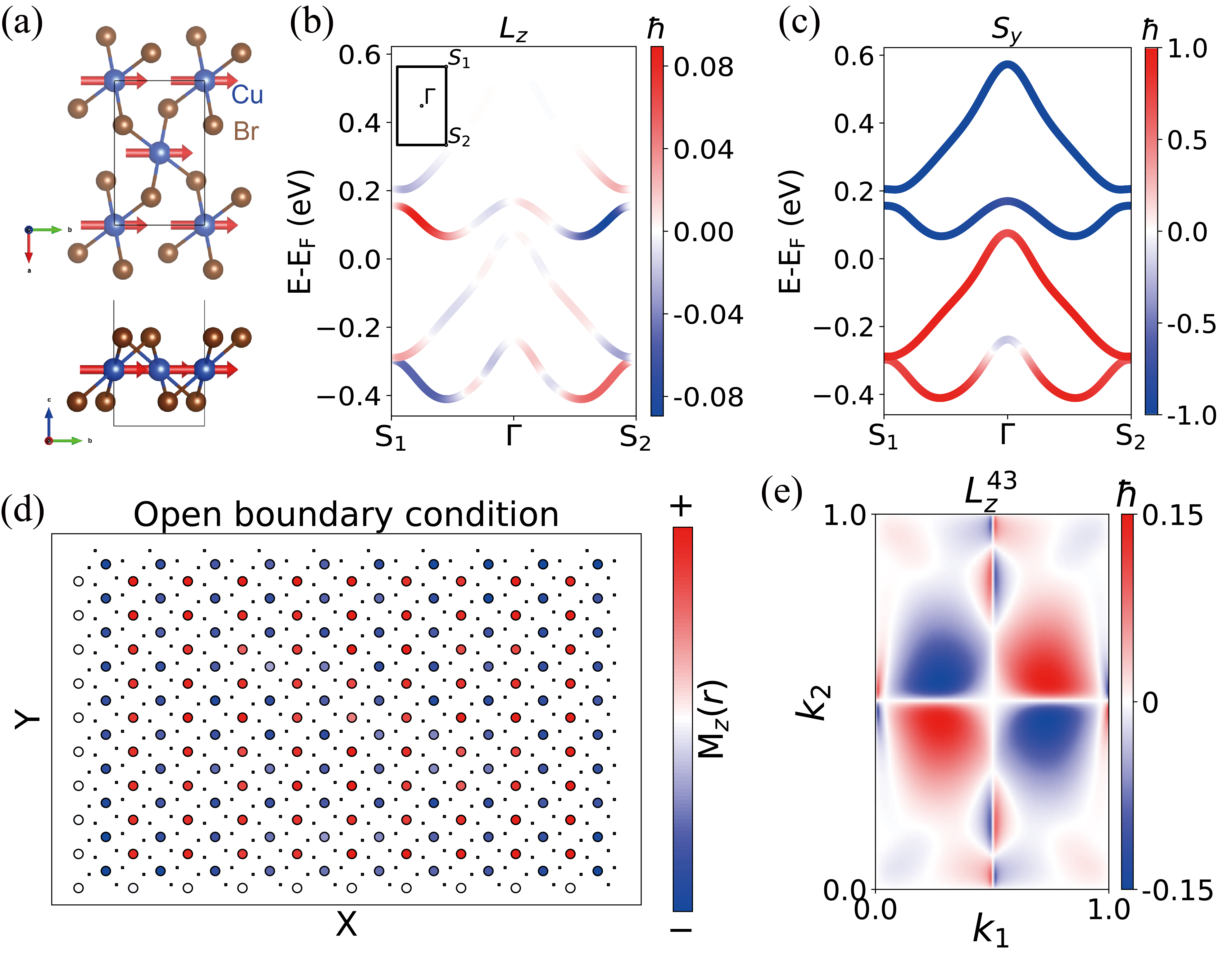}
\caption{\label{fig:cubr} (a) Top and side views of the monolayer CuBr$_2$. Spin magnetic moments localized on Cu atoms are annotated with red arrows. (b,c) Band structures of CuBr$_2$, with the color denoting the magnitude of $L_z$ (b) and $S_y$ (c) for each band. The inset displays the first Brillouin zone and high-symmetry points, with $S_1=(0.5,0.5)$ and $S_2=(0.5,-0.5)$. (d) Real-space distribution of orbital magnetic moments in a $10\times10$ lattice under open boundary conditions at $\mu=0$ eV. Red (blue) denotes positive (negative) $M_z(\rr)$. (e) Momentum-space distribution of $L_z(\kk)$ for the lowest conduction band (the 43rd band) across the Brillouin zone. }
\end{figure}

\textit{Orbital altermagnetism in spin ferromagnets.}---Strikingly, we find that orbital altermagnetism can exist in certain ferromagnets. Although the modern theory of orbital magnetization dictates that 2D systems lack in-plane orbital magnetic moments, the out-of-plane component $M_z$ can adopt an alternating arrangement even in the absence of staggered spin order. From a symmetry perspective, altermagnetism requires the magnetic point group to belong to the first or third magnetic class (systems with $\tau\mathcal{T}$ symmetry are excluded here). Notably, several of these magnetic groups are compatible with in-plane spin ferromagnetism. In a 2D ferromagnet with in-plane spins and belonging to these classes, alternating $M_z$ is allowed if the symmetry connects two sites with opposite $M_z$ but excludes $\mathcal{PT}$ and $C_{2z}\mathcal{T}$ symmetries. Only six magnetic point groups meet this criterion: $2$, $m$, $m_z^\prime$, $2/m$, $m_z^\prime m 2^\prime$, and $m^\prime m_z^\prime 2$.

In such \emph{orbital-altermagnetic ferromagnets}, the role of spin alignment is merely to break time-reversal symmetry, while the alternating $M_z$ pattern satisfies the magnetic point-group requirements. Two distinct cases arise: (1) Site-type orbital altermagnetism, where symmetry directly connects inequivalent atomic sites and reverses their $M_z$; and (2) Loop-type orbital altermagnetism, where site symmetry forces $M_z=0$ on all atomic sites, but the magnetic group permits circulating inter-site currents that generate staggered $M_z$ located at empty plaquette centers.

A prototypical site-type example is CuBr$_2$ \cite{adma.201200734}. We consider an in-plane ferromagnetic configuration [Fig.~\ref{fig:cubr}(a)], where the Cu spins align ferromagnetically along $y$, giving the magnetic point group $2/m$. The two Cu atoms are related by either a glide mirror $\widetilde{m}_y=\{m_y|\tfrac12,\tfrac12,0\}$ or a glide rotation $\widetilde{C}_{2y}=\{C_{2y}|\tfrac12,\tfrac12,0\}$, both of which reverse $M_z$. Our first-principles calculations with SOC indeed reveal opposite-sign $M_z(\rr)$ on two Cu atoms in real space, as shown in Fig.~\ref{fig:cubr}(d). Furthermore, the band structure along the diagonal and antidiagonal $\kk$-paths ($S_1$-$\Gamma$-$S_2$) exhibits opposite $L_z$ polarization with finite amplitude ($\sim 0.1\hbar$) [Fig.~\ref{fig:cubr}(b)], while maintaining persistent spin $S_y$ polarization [Fig.~\ref{fig:cubr}(c)]. The $\kk$-space distribution of $L_z(\kk)$ further manifests a defining characteristic $d$-wave pattern [see Fig.~\ref{fig:cubr}(e)].
Rotating the spins to the $x$ axis changes the magnetic group to $2'/m'$, where the $\widetilde{m}_y\mathcal{T}$ symmetry forces identical $M_z$ on both Cu atoms, thereby suppressing orbital altermagnetism. Consequently, the presence or absence of orbital altermagnetism for different spin alignments produces distinct nonlinear orbital Edelstein effects to be used as an experimental signature of orbital altermagnetism (see SM~\footnotemark[\value{footnote}]).

\begin{figure*}
\centering
\includegraphics[width=0.7\linewidth]{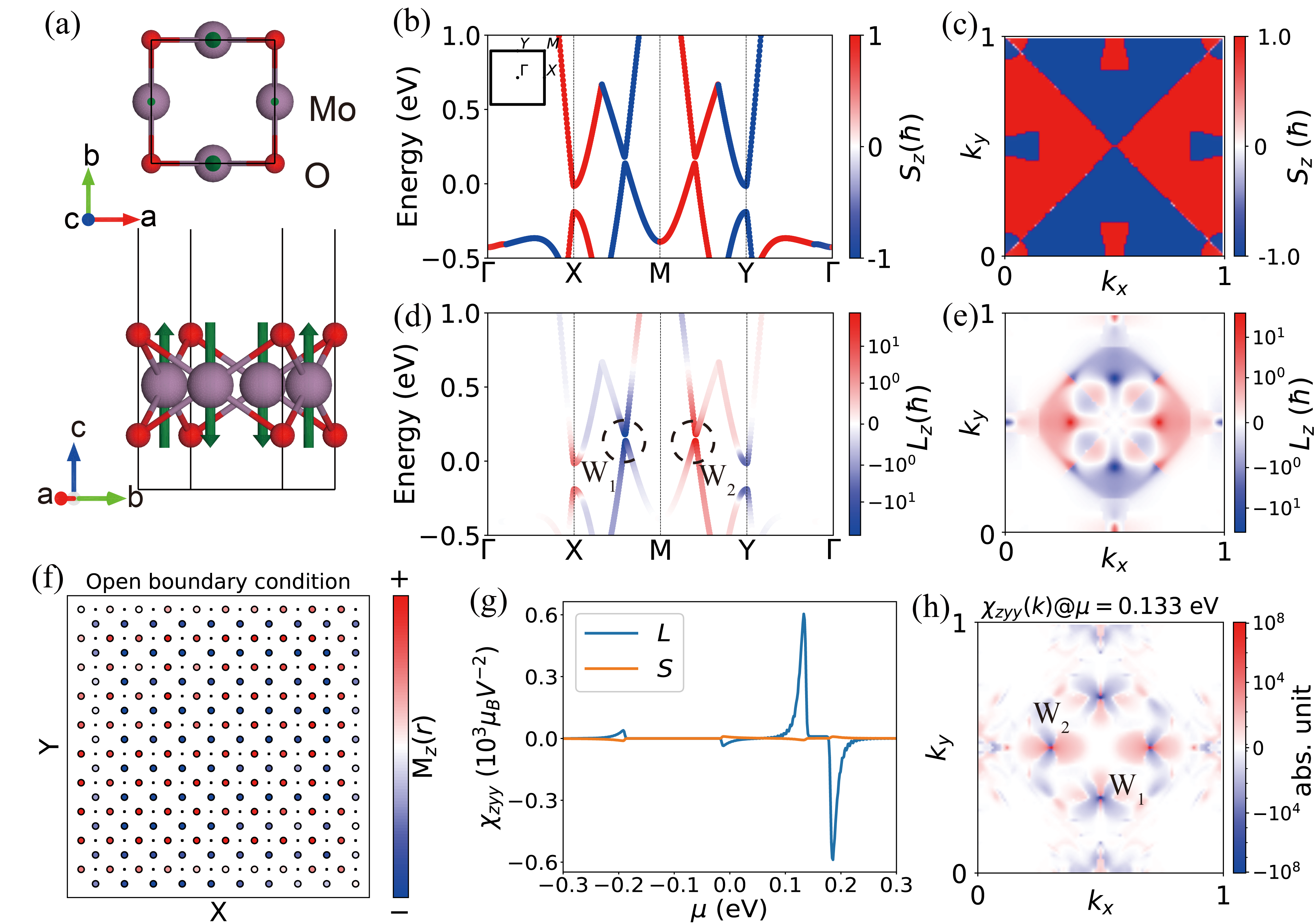}
\caption{\label{fig:MoO} (a) Top and side views of the monolayer MoO lattice. Green arrows indicate the spin magnetic moments on Mo atoms. (b) Spin-($S_z$) and (d) Orbital- ($L_z$) resolved band structures, where nearly degenerated band crossing points at $\mu=0.133$ eV are marked as $W_1$ and $W_2$. The inset displays the Brillouin zone with high-symmetry points. (c,e) Momentum-space distribution of $S_z(\kk)$ (c) and $L_z(\kk)$ (e) for the highest valence band across the Brillouin zone, with $k_x$ and $k_y$ in reduced coordinates. Note that the color maps of $L_z(\kk)$ are presented in logarithmic scale for better visualization of its $\bm{\kk}$-space variation.
(f) Real-space distribution of orbital magnetic moments in a $10\times 10$ lattice under open boundary conditions at $\mu$=0 eV. Red (blue) denotes positive (negative) $M_z(\rr)$. (g) Nonlinear responses $\chi^X_{zyy}$ ($X=L$ or $S$) as a function of chemical potential $\mu$ from first-principles calculations. (h) Momentum-space distribution of $\chi_{zyy}(\kk)$ across the Brillouin zone at $\mu=0.133$ eV.
}
\end{figure*}

In addition to the above site-type orbital altermagnetism, we identify the honeycomb VS$_2$ monolayer \cite{acs.nanolett.7b01914} as a prototypical loop-type orbital altermagnetic ferromagnet (see Sec.~VII in SM~\footnotemark[\value{footnote}]). When the spins align along the $a$ axis, the system belongs to the magnetic space group $2/m$, which forbids out-of-plane orbital magnetic moments at both V and S atomic sites. However, our calculations reveal two incoming currents of 0.44 $\mu$A and four outgoing currents of 0.22 $\mu$A on each V atom along the V-S bonds. These currents form closed loops within the V-S-V-S rhombi, circulating in opposite directions on adjacent plaquettes. This staggered loop-current pattern generates alternating out-of-plane orbital magnetic moments, thereby realizing orbital altermagnetism. The $L_z$-resolved band structure and the momentum-space distribution $L_z(\kk)$ (Fig.~S7 in SM~\footnotemark[\value{footnote}]) confirm the $d$-wave-like orbital-momentum locking,  characteristic of orbital altermagnetism.

\textit{Orbital altermagnetism in conventional spin altermagnets.}---We next illustrate the coexistence of spin and orbital altermagnetism using monolayer MoO as a representative example. The lattice structure [Fig.~\ref{fig:MoO}(a)] belongs to the $4^\prime/mm^\prime m$ magnetic point group. Its band structure exhibits clear momentum-dependent $S_z$ and $L_z$ polarizations, indicating simultaneous spin and orbital altermagnetic order [Fig.~\ref{fig:MoO}(b,d)]. The alternating momentum-space distribution of $L_z(\kk)$ for the highest valence band [Fig.~\ref{fig:MoO}(c,e)] and the staggered real-space arrangement of orbital magnetic moments [Fig.~\ref{fig:MoO}(f)] confirm a $d$-wave-type orbital altermagnetism. Similar behavior is also found in monolayer CrO \cite{npjCompMat2023CrO}, with results presented in SM~\footnotemark[\value{footnote}].

Like its spin counterpart, orbital altermagnetism contributes to a variety of magnetic phenomena, including the magneto-optical Kerr effect~\cite{Iguchi_2025,10.1063/5.0244878}, X-ray magnetic dichroism~\cite{PhysRevLett.132.176701,cm39-hxqk}, piezomagnetism~\cite{JunweiLiu2021Natcomm, PhysRevX.15.021083}, and nonlinear Edelstein effect~\cite{PhysRevLett.129.086602, PhysRevLett.130.166302,baek_nonlinear_2024}.
For instance, in the nonlinear Edelstein effect, a nonlinear current-induced spin or orbital magnetization can be expressed as $\delta X_a=\chi^X_{abc}E_bE_c$, where $a, b, c\in \{x, y, z\}$, $E_a$ is the electric field, and $X=S\, \mathrm{or}\, L$ denote the spin or orbital contributions, respectively. The nonlinear response tensor $\chi^X_{abc}$ is governed by band-geometric quantities such as the quantum metric, Berry curvature, and the matrix elements of spin or orbital angular momentum (see Sec. III in SM for details~\footnotemark[\value{footnote}]). In monolayer MoO, the combined symmetries $C_{4z}\mathcal{T}$ and $m_x\mathcal{T}$ strongly constrain the second-order response tensor $\chi^X_{z i j}$. For the out-of-plane response ($z$ index), the time-reversal-even part is symmetry-forbidden, leaving only the intrinsic (time-reversal-odd) contribution. In this case, $m_x\mathcal{T}$ enforces $\chi^X_{zxy}=0$, while $C_{4z}\mathcal{T}$ imposes $\chi^X_{zxx}=-\chi^X_{zyy}$. Consequently, for an in-plane electric field $\mathbf{E}=E(\sin\theta,\cos\theta)$, the induced out-of-plane polarization varies as
$\delta X_z\propto E^2\cos 2\theta$, leading to a characteristic $\pi$-periodic angular dependence.

Remarkably, as shown in Fig.~\ref{fig:MoO}(g), the orbital contribution $\chi^L_{zyy}$ is substantially larger than its spin counterpart $\chi^S_{zyy}$, indicating that the current-induced magnetization is dominated by orbital magnetization. A pronounced peak in $\chi^L_{zyy}$ occurs at $\mu = 0.133$ eV, originating from prominent orbital magnetic moments near the nearly degenerate band crossing points ($W_1$ and $W_2$) along the $X$-$\Gamma$ and $Y$-$\Gamma$ paths [see Fig.~\ref{fig:MoO}(d)] \cite{PhysRevB.111.075145, pnas.2305541120}. The $\kk$-space distribution of $\chi^L_{zyy}(\kk)$ clearly shows that significant contributions are concentrated around the nearly degenerated points [Fig.~\ref{fig:MoO}(h) and Fig.~S4 and S5 in SM~\footnotemark[\value{footnote}]].
Since these points can be tuned close to the Fermi level via electric gating or tensile strain~\cite{PhysRevB.107.214419}, we anticipate that a large nonlinear Edelstein response could be experimentally accessed through gating or strain, serving as a clear signature of orbital altermagnetism in monolayer MoO.

\textit{Discussion and summary}---Recent studies have shown that weak orbital ferromagnetism or ferrimagnetism can emerge in altermagnets ~\cite{sciadv.aaz8809, bozorth1958ferromagnetic, PhysRevB.100.224425, PhysRevLett.134.196703} via SOC, arising from alternating $g$-tensor anisotropy or noncollinear magnetic moments. Such states are classified as M-type altermagnets~\cite{Cheong2025Alterm_classification, Cheong2024alterm_noncollinear}, which—in absence of SOC—exhibit perfect spin compensation, but develop a finite net magnetization from orbital magnetization and spin canting in the presence of SOC, and belong to ferromagnetic point groups.
In contrast, our work reveals an inversed scenario: systems where orbital magnetic moments display a symmetry-protected altermagnetic order, yet the spin sector remains uncompensated (finite net spin). This concept naturally extends to spin ferrimagnets and noncollinear magnets belonging to ferromagnetic point groups. Moreover, the orbital altermagnetic state can, in principle, be tuned by spin reorientation (which changes the magnetic group symmetry) or by electric fields via orbital magnetoelectric coupling, offering new avenues for controlling altermagnetic responses. Recent loop-current-driven \cite{leeb2026collinearpwavemagnetismhidden} and orbital-order-driven \cite{PhysRevLett.132.236701,cjzw-j4v7} altermagnetic states are conceptually related but distinct from our work. There, the orbital sector mainly breaks additional symmetries so as to enable or modify spin altermagnetism. Here, by contrast, the orbital degree of freedom itself forms the altermagnetic texture, while the spin sector mainly plays a symmetry-setting and tuning role.

Overall, our findings identify orbital altermagnetism as a broad and robust class of magnetic order---one that is distinct from, yet intertwined with, spin altermagnetism---opening a pathway to novel symmetry-driven magnetotransport phenomena and electrically tunable orbital-based spintronic devices. Experimentally, the significant orbital magnetization establishes giant current-induced orbital polarization as a transport signature of orbital altermagnetism and suggests that orbital degrees of freedom may play a dominant role in certain magnetic responses. We expect our theoretical proposal will stimulate immediate experimental interests, since recent experimental developments have made the detection of the real-space orbital magnetization patterns and altermagnetic orbital texture in momentum space possible, such as circular dichroism combined with soft x-ray angle-resolved photoemission spectroscopy~\cite{PhysRevLett.132.196402}, and nitrogen-vacancy (NV)-center magnetometry~\cite{NVcenter2025Nagaosa} (see Sec. VIII in SM~\footnotemark[\value{footnote}] for detailed discussion).
In addition to conventional magnetic materials studied here, the orbital altermagnetism is also expected to be achieved from the orbital antiferromagnetism in correlated systems, such as cuprates and cerium hexaboride CeB$_6$, via various symmetry-engineering approaches such as strain\cite{PhysRevB.109.144421}, electric field~\cite{cxvd-mp2h, mazin2023inducedmonolayeraltermagnetismmnpsse3, shvd-vmvs}, controllable stacking bilayers \cite{PhysRevLett.130.046401, PhysRevLett.133.166701, PhysRevLett.133.206702, PhysRevMaterials.8.L051401}, ferroelectric substrate \cite{acs.nanolett.4c02248}, and Janusization \cite{shvd-vmvs, acs.nanolett.5c00198}.

\begin{acknowledgments}
This work is supported by the National Key R\&D Program of China (Grant No. 2021YFA1401600) and the National Natural Science Foundation of China (Grants No. 12074006 and 12474056). F.L. acknowledges support from DOE-BES (Grant No. DEFG02-04ER46148). The work was carried out at the National Supercomputer Center in Tianjin, and the calculations were performed on Tianhe new generation supercomputer. The high-performance computing platform of Peking University supported the computational resources.
\end{acknowledgments}

\textit{Data availability}---The data that support the findings of this article are openly available~\cite{huang_2026_19602057}.

\bibliography{apssamp}

\begin{widetext}

\appendix

\section{Formula of Orbital Magnetization in Reciprocal and Real Spaces}
Orbital magnetism arises from the motion of electrons in atomic orbitals or Bloch states and plays a crucial role in diverse magnetic and transport phenomena, particularly in systems where spin contributions are suppressed or symmetry-forbidden. Unlike spin magnetic moments---typically described by exchange interactions or Zeeman coupling---orbital magnetic moments originate from the circulation of charge in real or momentum space and are intrinsically tied to the geometric and topological properties of electronic wavefunctions.

From a theoretical standpoint, the orbital magnetic moment of a Bloch electron in band $n$ at wavevector $\kk$ can be expressed as the expectation value of the self-rotation of a wave packet about its center of mass. This is formalized in the modern theory of orbital magnetization, which extends semiclassical concepts into a gauge-invariant quantum framework. In two dimensions, only the out-of-plane component is well defined. At zero temperature in a two-dimensional (2D) periodic crystal, the total orbital magnetization is given by \cite{PhysRevLett.95.137205, PhysRevB.74.024408, cphc.200400641, PhysRevB.106.075136}:
\begin{equation}
\begin{split}
M &= M_{\textnormal{LC}} + M_{\textnormal{IC}} + M_{\textnormal{BC}},\\
M_{\textnormal{LC}} &= \frac{e}{2\hbar c}  \textnormal{Im} \sum_n \int_{\epsilon_{n\kk} \le \mu} [d\kk]
\bra{\partial_{\kk} u_{n\kk}} \times H_{\kk} \ket{\partial_{\kk} u_{n\kk}}, \\
M_{\textnormal{IC}} &= \frac{e}{2\hbar c}  \textnormal{Im} \sum_n \int_{\epsilon_{n\kk} \le \mu} [d\kk]
\epsilon_{n\kk}  \bra{\partial_{\kk} u_{n\kk}} \times \ket{\partial_{\kk} u_{n\kk}}, \\
M_{\textnormal{BC}} &= -\mu \frac{e}{2\pi \hbar c} \mathcal{C}, \\
\mathcal{C} &= \frac{1}{2\pi} \textnormal{Im} \sum_n \int_{\epsilon_{n\kk} \le \mu} d\kk
\bra{\partial_{\kk} u_{n\kk}} \times \ket{\partial_{\kk} u_{n\kk}},
\end{split}\label{M_LC_IC_BC}
\end{equation}
where $|u_{n\kk}\rangle$ is the cell-periodic Bloch function and $\epsilon_{n\kk}$ the band energy. All summations in Eq.~\eqref{M_LC_IC_BC} are over occupied bands $n$ up to $\mu$. Here, $M_{\textnormal{LC}}$ corresponds to the local circulation (LC) of Wannier orbitals, $M_{\textnormal{IC}}$ arises from the itinerant circulation (IC) of Bloch electrons, and $M_{\textnormal{BC}}$ is a topological contribution proportional to the Chern number $\mathcal{C}$ (quantized when $\mu$ is in the bulk gap), which is finite only in systems with broken time-reversal symmetry and nontrivial topology.

To resolve orbital-momentum locking in momentum space for 2D systems with trivial topology ($\mathcal{C}=0$), we employ the band- and $\kk$-resolved expression of the orbital magnetic moment \cite{Resta_2010, OM_in_solids, RevModPhys.82.1959, PhysRevB.74.024408},
\begin{equation}
{M}_{n\kk}=\frac{e}{2\hbar c} \textnormal{Im}
\bra{{\partial}_\kk u_{n \kk}} \times
\left(H_\kk-\epsilon_{n\kk}\right)\ket{{\partial}_\kk u_{n\kk}},
\label{eq:m-orb}
\end{equation}
More specifically,
\begin{equation}
M_{n\kk} = -\frac{e}{2\hbar c}\,\mathrm{Im}\!\sum_{m\neq n}
\frac{\langle u_{n\kk}|\hat{\mathbf{v}}|u_{m\kk}\rangle \times
\langle u_{m\kk}|\hat{\mathbf{v}}|u_{n\kk}\rangle}
{\varepsilon_{n\kk}-\varepsilon_{m\kk}},
\end{equation}
where $\hat{\mathbf{v}} = \frac{1}{\hbar}\nabla_{\kk} H(\kk)$ is the velocity operator. From this expression, the orbital magnetic moment is expected to increase significantly in regions where the bands are nearly degenerate.
In two dimensions, $M_{n\kk}$ is along the out-of-plane direction and is directly related to the orbital angular momentum via
\begin{equation}
M_{n\kk} = - g_L \mu_B L_{z}(\kk),\label{Mz_Lz}
\end{equation}
where $\mu_B$ is the Bohr magneton and $g_L=1$ the Land\'e g-factor \cite{PhysRevB.103.195309}.

To characterize orbital magnetization in systems lacking full translational symmetry (e.g., under open boundary conditions or disorder), a real-space formulation is often more appropriate. Analogous to the above $\kk$-space integration, orbital magnetization and its constituent terms can also be expressed as a real-space integration $M=\frac{1}{A}\int_Ad\rr m(\rr)$, which has the form of a real-space average over some region with an area $A$. Following Refs.~\cite{PhysRevLett.110.087202, PhysRevB.95.121114, PhysRevB.106.075136}, the corresponding orbital magnetization density $m(\rr)$ can be written as a sum of gauge-invariant local markers:
\begin{equation}
\label{eq:realmarker}
\begin{split}
    m(\rr) &= m_{\textnormal{LC}}(\rr) + m_{\textnormal{IC}}(\rr) + m_{\textnormal{BC}}(\rr), \\
    m_{\textnormal{LC}}(\rr) &= \frac{e}{\hbar c}\textnormal{Im} \bra{\rr} P x Q H Q y P \ket{\rr}, \\
    m_{\textnormal{IC}}(\rr) &= -\frac{e}{\hbar c}\textnormal{Im} \bra{\rr} Q x P H P y Q \ket{\rr}, \\
    m_{\textnormal{BC}}(\rr) &= -\mu \frac{2e}{\hbar c}  \textnormal{Im} \bra{\rr} Q x P y Q \ket{\rr},
\end{split}
\end{equation}
where $P$ is the ground-state projector and $Q=1-P$ its complement:
\begin{equation}
P = \sum_{\epsilon_i \le \mu} \ket{\varphi_i} \bra{\varphi_i}, \qquad H \ket{\varphi_i} = \epsilon_i \ket{\varphi_i}.
\end{equation}
This formalism provides a position-resolved, gauge-invariant description of orbital magnetization, enabling direct evaluation of orbital textures in both tight-binding and first-principles frameworks.

In the main text, we apply this approach to the minimal tight-binding model of orbital altermagnetism. The resulting real-space orbital magnetization $m(\rr)$, computed from Eq.~(\ref{eq:realmarker}), serves as a key diagnostic for identifying staggered orbital ordering patterns in real space.

\section{Magnetic octupole and Magnetic Toroidal quadrupole}
In both theoretical models and real magnetic materials, orbital magnetic moments can often be interpreted as arising from local loop currents. Accordingly, the magnetic (M) multipoles and magnetic toroidal (MT) multipoles are defined from the explicit current distribution. Previous studies~\cite{doi:10.7566/JPSJ.93.072001} have established that imaginary hopping in tight-binding models corresponds to MT multipoles, thereby linking orbital magnetism intrinsically to these multipoles.

Magnetic multipoles $M_{lm}$ and magnetic toroidal multipoles $T_{lm}$ exhibit distinct transformation properties under spatial inversion $\mathcal{P}$ and time-reversal symmetry $\mathcal{T}$ \cite{PhysRevB.98.165110}:
\[
    M_{lm} \ \xrightarrow{(\mathcal{P,T})} \ [(-1)^{l+1}, -1], \quad
    T_{lm} \ \xrightarrow{(\mathcal{P,T})} \ [(-1)^{l}, -1].
\]
From these transformation rules, it follows that collinear altermagnets can uniquely host MT quadrupoles or M octupoles, which are absent in conventional ferromagnets or antiferromagnets. 

The explicit definitions are given as~\cite{doi:10.7566/JPSJ.87.033709}:
\begin{equation}
    M_{lm} = \frac{1}{c(l+1)} \int dr \, [\mathbf{r} \times \mathbf{j}_e(\mathbf{r})] \cdot \nabla O_{lm}(\mathbf{r}),
\end{equation}
\begin{equation}
    T_{lm} = \frac{1}{c(l+1)} \int dr \, [\mathbf{r} \cdot \mathbf{j}_e(\mathbf{r})] \, O_{lm}(\mathbf{r}),
\end{equation}
where $\mathbf{j}_e(\mathbf{r})$ is the electric current density, and
\[
    O_{lm}(\mathbf{r}) = \sqrt{\frac{4\pi}{2l+1}} \, r^l Y^*_{lm}(\hat{\mathbf{r}})
\]
is expressed in terms of the spherical harmonics $Y_{lm}(\hat{\mathbf{r}})$, with $l$ and $m$ denoting the azimuthal and magnetic quantum numbers, respectively.

For the minimal model defined in Eq.~(1) in the main text, we reveal that $T_{xy}\neq0$ and $M_{z}^{\beta}\neq 0$ (with $O_z^\beta=\frac{\sqrt{15}}{2}z(x^2-y^2))$, consistent with the allowed multipoles of the corresponding magnetic point group \cite{PhysRevB.104.054412}. Particularly, the nonzero magnetic octupole $M_{z}^{\beta}$ can serve as a secondary order parameter in the Landau theory of the $d$-wave altermagnetism \cite{PhysRevLett.132.176702}. 
The nonzero $T_{xy}$ can also be directly traced back to the imaginary hopping terms in the tight-binding model. By evaluating the symmetry-adapted multipole basis (SAMB) for the bond cluster \cite{PhysRevB.107.195118}, we find that in the D$_{4\mathrm{h}}$ point group, only the irreducible representation B$_{2\mathrm{g}}$ yields a non-zero toroidal quadrupole, which precisely corresponds to $T_{xy}$.

\section{Intrinsic nonlinear current-induced spin and orbital magnetization}
In this section, we provide a brief introduction to the formalism of the intrinsic nonlinear Edelstein effect.
For a 2D system with either $C_{2z}$ or $\mathcal{P}$ symmetry, the linear spin and orbital responses are forbidden by symmetry. In such cases, the leading contribution arises at the second order:
\begin{equation}
    \delta X_a=\chi^X_{abc}E_bE_c,
\end{equation}
where $X_a$ denotes the $a$-th component of either the orbital angular momentum $L_a$ or the spin angular momentum $S_a$. The nonlinear response tensor $\chi_{abc}$ can be decomposed according to its parity under time-reversal symmetry \cite{PhysRevLett.129.086602,PhysRevLett.130.166302,baek_nonlinear_2024,dww3-vm74,4kmy-59l9}:
\begin{equation}
\begin{split}
    \chi_{abc}^\mathrm{even}&=\tau\sum_n\int[d\kk]f_0 \partial_c\gamma^{X,n}_{ab},\\
    \chi_{abc}^\mathrm{odd}&=\frac{1}{2}\int[d\kk]\sum_n\Bigg[ \Lambda_{abc,n}^Xf_0-(X_{a,n}G_{bc,n}-v_{b,n}\mathscr{G}^X_{ac,n}-v_{c,n}\mathscr{G}^X_{ab,n})f^\prime_0\Bigg].
\end{split}
\end{equation}
Here $[d\kk]\equiv d\kk / (2\pi)^d$ is a shorthand notation for integration over the Brillouin zone in $d$ dimensions, and $f_0$ is the Fermi-Dirac distribution with unperturbed band energy $\epsilon_n$, and $f^\prime_0$ is its derivative. The quantity $X_{a,n}=\langle u_n|\hat{X}_a|u_n\rangle$ is the intraband matrix element of spin or orbital angular momentum, and $v_{a,n}$ is the intraband velocity matrix element.

The Zeeman-like field $\boldsymbol{h}$ couples to the spin or orbital degree of freedom in the form of $- \boldsymbol{\hat{s}} \cdot \boldsymbol{h}$ or $-\boldsymbol{\hat{L}} \cdot \boldsymbol{h}$. Therefore, in the presence of an external field, one can define the Berry connection and the related quantities in the $h$-space \cite{PhysRevLett.129.086602}. For example, the $h$-space Berry connection is defined as $\mathfrak{A}_i=\langle u_n(\kk)|i\partial_{h_i}|u_n(\kk)\rangle$. The Berry connection polarizability (BCP) and the $h$-space BCP are defined as
\begin{equation}
\begin{split}
    G_{ab,n}&=2\textnormal{Re} \sum_{m\neq n}\frac{v_{a,nm}v_{b,mn}}{(\epsilon_n-\epsilon_m)^3},\\
    \mathscr{G}^S_{ab,n}&=-2\textnormal{Re} \sum_{m\neq n}\frac{s_{a,nm}v_{b,mn}}{(\epsilon_n-\epsilon_m)^3},\\
    \mathscr{G}^O_{ab,n}&=-2\textnormal{Re} \sum_{m\neq n}\frac{M_{a,nm}v_{b,mn}}{(\epsilon_n-\epsilon_m)^3},
\end{split}
\end{equation}
where the numerator involves the interband matrix elements of spin and velocity operators, $s_{a,nm}$ and $v_{a,nm}$, $M_{mn}=\sum_{l\neq n}(v_{ml}+\delta_{lm}v_{nn})\times A_{ln}/2$ is the interband orbital magnetic moment, $A_{mn}=\langle u_m|i\partial_\kk|u_n\rangle$ is the Berry connection.
The anomalous spin polarizability (ASP) and the anomalous orbital polarizability (AOP) is given by
\begin{equation}
\begin{split}
    \gamma^{S,n}_{ab}(\kk)&=2g\mu_B\textnormal{Im}\sum_{m\neq n}\frac{s_{a,nm}v_{b,mn}}{(\epsilon_n-\epsilon_m)^2},\\
    \gamma^{L,n}_{ab}(\kk)&=-2\textnormal{Im}\sum_{m\neq n}\frac{M_{a,nm}v_{b,mn}}{(\epsilon_n-\epsilon_m)^2}-\frac{e}{2\hbar}\epsilon_{akl}\Gamma_n^{lkb},
\end{split}
\end{equation}
where the quantum Christoffel symbol is defined as
\begin{equation}
\begin{split}
    \Gamma_n^{abc}&=\frac{1}{2}(\partial_cg_n^{ba}+\partial_bg_n^{ac}-\partial_ag_n^{bc}),\\
    g_n^{ab}&=\mathrm{Re}\sum_{m\neq n}A_{nm}^iA_{mn}^j.
\end{split}
\end{equation}
The second-order response kernel $\Lambda_{abc,n}^X$ is given by
\begin{equation}
\begin{split}
    \Lambda_{abc,n}^X=-2\textnormal{Re}&\sum_{m\neq n}\Bigg[\frac{3v_{b,nm}v_{c,mn}(X_{a,nn}-X_{a,mm})}{(\epsilon_n-\epsilon_m)^4}
    -\frac{(\partial_bX_a)_{nm}v_{c,mn}+(\partial_cX_a)_{nm}v_{b,mn}}{(\epsilon_n-\epsilon_m)^3}\\
    -&\sum_{l\neq n}\frac{(v_{b,lm}v_{c,mn}+v_{c,lm}v_{b,mn})X_{a,nl}}{(\epsilon_n-\epsilon_l)(\epsilon_n-\epsilon_m)^3}
    -\sum_{l\neq m}\frac{(v_{b,ln}v_{c,nm}+v_{c,ln}v_{b,nm})X_{a,ml}}{(\epsilon_m-\epsilon_l)(\epsilon_n-\epsilon_m)^3}
    \Bigg].
\end{split}
\end{equation}
Finally, the orbital angular momentum (OAM) matrix elements follow from the modern theory of orbital magnetization \cite{PhysRevB.106.104414}:
\begin{equation}
\begin{split}
    \langle u_n|\hat{\boldsymbol{L}}|u_m\rangle&=i\frac{e\hbar^2}{4g_L\mu_B}\sum_{q\neq n,m}(\frac{1}{\epsilon_q-\epsilon_n}+\frac{1}{\epsilon_q-\epsilon_m})\cdot\langle u_n|\hat{\boldsymbol{v}}|u_q\rangle\times\langle u_q|\hat{\boldsymbol{v}}|u_m\rangle.\label{Lz_k}
\end{split}
\end{equation}
The diagonal element of the orbital angular momentum $\bm{L}_n$ is linearly related to the orbital magnetization. In 2D systems, its z-component corresponds to the out-of-plane component of the orbital magnetic moment via Eq.~\eqref{Mz_Lz} \cite{PhysRevB.103.195309}.

Note that the matrix elements of the orbital angular momentum (OAM) here are different from those of the atom-centered approximation (ACA), which assumes that the OAM originates from the region centered on the atomic sites. Within ACA, $\hat{\boldsymbol{L}}$ reduces to the on-site atomic orbital angular momentum operator that mixes orbitals located at the same lattice site~\cite{PhysRevB.106.104414, PhysRevResearch.5.043052}. While the ACA is often adequate for wide-gap semiconductors, it is known to fail in both narrow-gap systems and in transition metals.

\section{Comparison between the modern theory and the atom-centered approximation for orbital angular momentum}
\label{sec:modern_vs_aca}

In this section, we briefly compare the orbital angular momentum obtained from the modern theory of orbital magnetization and that obtained within the atom-centered approximation (ACA). Since the modern-theory expression has already been introduced in the previous section, here we focus on the ACA description and its relation to the momentum-space orbital pattern discussed in the main text.

Within the ACA, the orbital angular momentum is evaluated by projecting the Bloch eigenstates onto local atomic orbitals and then calculating the expectation value of the corresponding orbital angular momentum operator. For a Bloch state $\ket{\psi_{n\mathbf{k}}}$, the ACA orbital angular momentum is written as
\begin{equation}
\mathbf{L}^{\mathrm{ACA}}_{n}(\mathbf{k})
=
\bra{\psi_{n\mathbf{k}}}\hat{\mathbf{L}}^{\mathrm{loc}}\ket{\psi_{n\mathbf{k}}},
\end{equation}
where $\hat{\mathbf{L}}^{\mathrm{loc}}$ denotes the local orbital angular momentum operator defined in the Wannier-orbital basis. In this approximation, the orbital angular momentum is viewed as a sum of atom-centered local contributions, and therefore it captures the atomic-like part of the orbital character of the Bloch states.

Compared with the modern theory, the ACA does not explicitly include the itinerant intercell contribution associated with the geometric structure of Bloch wave functions. Therefore, the ACA is in general not expected to reproduce the full quantitative value of the orbital angular momentum obtained from the modern theory, especially in the presence of strong interband effects near near-degenerate regions. Nevertheless, as we show below, once the symmetry of the system is properly respected, the ACA can still reproduce the same symmetry-determined orbital altermagnetic pattern in momentum space. The main differences between the two approaches lie in the quantitative magnitude and the finer details of the distribution. In addition, within the ACA the orbital angular momentum can in general have components along different spatial directions, whereas for the two-dimensional modern-theory formula used in the present work only the out-of-plane component is relevant.

As a representative example, we consider monolayer MoO. In our Wannier basis, the low-energy electronic structure is mainly described by the Mo $d$ orbitals and O $p$ orbitals. For the purpose of evaluating the ACA orbital angular momentum, one needs the matrix representation of the $L_z$ operator in these orbital subspaces.

For the real $p$-orbital basis
\begin{equation}
\left( p_x,\; p_y,\; p_z \right),
\end{equation}
the matrix representation of $\hat{L}_z$ is
\begin{equation}
L_z^{(p)}
=
\hbar
\begin{pmatrix}
0 & -i & 0 \\
i & 0 & 0 \\
0 & 0 & 0
\end{pmatrix}.
\end{equation}

For the real $d$-orbital basis
\begin{equation}
\left( d_{z^2},\; d_{xz},\; d_{yz},\; d_{x^2-y^2},\; d_{xy} \right),
\end{equation}
the matrix representation of $\hat{L}_z$ is
\begin{equation}
L_z^{(d)}
=
\hbar
\begin{pmatrix}
0 & 0 & 0 & 0 & 0 \\
0 & 0 & -i & 0 & 0 \\
0 & i & 0 & 0 & 0 \\
0 & 0 & 0 & 0 & 2i \\
0 & 0 & 0 & -2i & 0
\end{pmatrix}.
\end{equation}

Using these matrices, we evaluate the ACA orbital angular momentum for MoO over the full Brillouin zone. As shown in Fig.~\ref{fig:acafig}(b), the resulting $L_z^{\mathrm{ACA}}(\mathbf{k})$ distribution exhibits the same even-parity $d$-wave-like orbital altermagnetic pattern as that obtained from the modern theory. This confirms that the momentum-space pattern is mainly constrained by symmetry.

Therefore, the comparison between the modern theory and the ACA suggests the following picture. The modern theory is necessary for a complete and quantitatively reliable description of the orbital angular momentum, especially when the orbital response is strongly enhanced by interband effects. In contrast, the ACA provides a simpler local-orbital description and, when the symmetry of the model is properly enforced, the resulting \(L_z(\kk)\) exhibits the same symmetry-characterized momentum-space pattern as in the modern-theory result. Nevertheless, the ACA orbital angular momentum is, in general, a three-dimensional vector and can have nonzero components other than \(L_z\). Therefore, the ACA does not imply that the orbital angular momentum is restricted to be purely out of plane at each \(\kk\). In this sense, although the \(L_z\) pattern remains symmetry-consistent, the full orbital-angular-momentum texture in the ACA cannot be inferred solely from \(L_z\), and the question of whether the orbital angular momentum is collinearly compensated cannot be judged from \(L_z\) alone.

Using this on-site Wannier \(\hat{L}_z\) operator, we further recompute the corresponding nonlinear response tensor as functions of chemical potential within the ACA, as shown in Fig.~\ref{fig:acafig}(c). In this on-site evaluation, the orbital angular momentum is typically of order \(0.1\hbar\), and the resulting nonlinear response is correspondingly much smaller than that obtained from the modern-theory formula, becoming comparable in magnitude to the spin-only contribution.

This explicit comparison further shows that the very large orbital-angular-momentum values and the associated strong nonlinear response reported in the main text do not originate from a trivial normalization factor or from the bare atomic \(p\)- or \(d\)-orbital eigenvalues. Instead, they mainly arise from the itinerant interband contribution captured by the modern theory of orbital magnetization, which can be strongly enhanced near band crossings or quasi-degenerate regions. Therefore, while the ACA can reproduce the same symmetry-characterized \(L_z(\kk)\) pattern when the symmetry is properly enforced, it substantially underestimates the magnitude of both the orbital angular momentum and the resulting response tensor in the present system.

\begin{figure}
\centering
\includegraphics[width=1.0\linewidth]{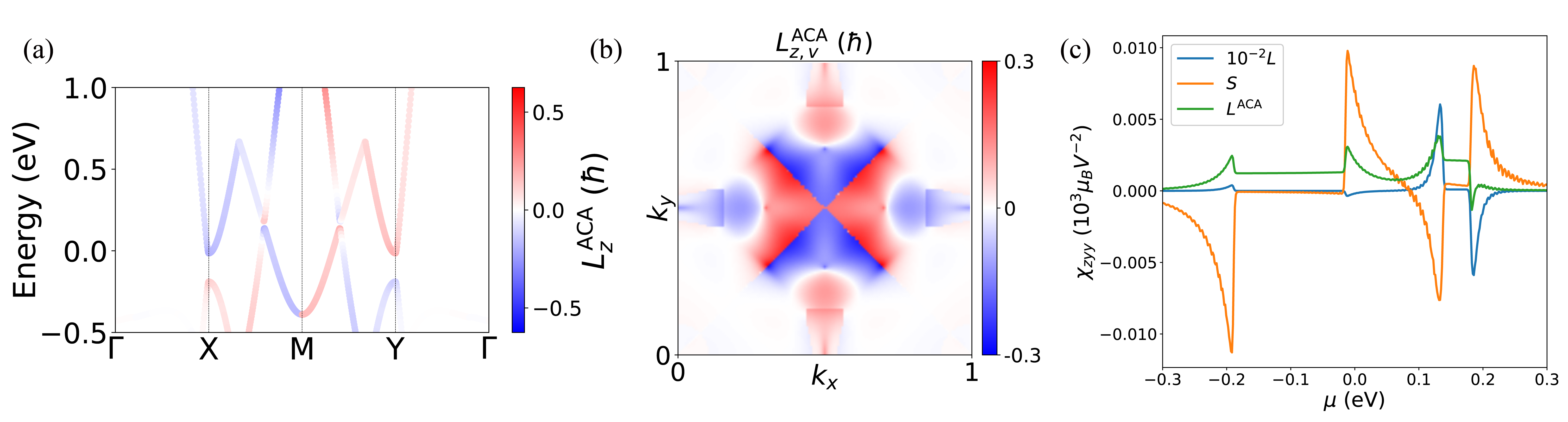}
\caption{\label{fig:acafig} (a) Band structures of MoO, with color indicating the magnitude of $L_z$ for each band and k point, calculated using the atom-centered approximation with $p$- and $d$-orbitals corresponding to the angular momentum. (b) The distribution of $L_z(\kk)$ for the highest valence band across the full Brillouin zone in MoO. (c) Nonlinear responses $\chi_{zyy}^X$ ($X=L$, $S$, or $L^{\mathrm{ACA}}$ ) as functions of chemical potential $\mu$. Here $L$ and $S$ represent the orbital and spin contributions, respectively, while $L^{\mathrm{ACA}}$  corresponds to the orbital angular momentum calculated using the ACA approach.}
\end{figure}

\section{Symmetry Analysis of Orbital Altermagnetism in Two Dimensions}

\subsection{Definition and symmetry constraints of 2D orbital altermagnetism}
We define a 2D orbital altermagnet as a system whose real space can be partitioned into two sublattices carrying nonzero but opposite orbital magnetic moments. The sublattices may consist either of atomic sites or of loop currents on bonds. In both cases, we focus on the out-of-plane component $M_z(\rr)$, which is the natural component in a two-dimensional geometry.

A key feature of orbital altermagnetism is that the net out-of-plane orbital magnetization vanishes, which the local distribution is not identically zero. This is the orbital analogue of a compensated antiferromagnetic-like or altermagnetic-like pattern, with the magnetic degree of freedom originating from orbital motion rather than from spin.

In momentum space, for an orbital altermagnet, the momentum-resolved orbital magnetic moment or the orbital angular momentum $L_z(\kk)$ must not be forced to vanish by symmetry across the Brillouin zone. In particular, any antiunitary symmetry that maps $\kk\to\kk$ while flipping the sign of the axial out-of-plane moment would enforce $L_z(\kk)=0$, and is therefore incompatible with a finite orbital moment texture. This includes, for example, $\mathcal{PT}$ or $C_{2z}\mathcal{T}$.

\subsection{Screening magnetic point groups compatible with 2D orbital altermagnetism}

We performed an exhaustive symmetry screening over all magnetic point groups to identify those that can host 2D orbital altermagnetism. The screening procedure follows directly from the definition and the symmetry constraints discussed above.

\paragraph*{(i) Remark on gray magnetic point groups and the underlying MSG type.
A gray (type-II) magnetic point group contains the pure time-reversal operation $\mathcal{T}$ at the point-group level.
Importantly, the \emph{same} gray point group can originate from two physically different situations at the magnetic space-group (MSG) level:
a type-II (gray) MSG, where $\mathcal{T}$ appears as $\{1|\mathbf{0}\}\mathcal{T}$ and the crystal is nonmagnetic; or a type-IV MSG, where the relevant antiunitary symmetry is a \emph{generalized} time reversal of the form
\begin{equation}
\tilde{\mathcal{T}}=\{1|\boldsymbol{\tau}\}\mathcal{T},
\end{equation}
with a nontrivial translation $\boldsymbol{\tau}$, which projects to $\mathcal{T}$ in the point-group description.}

Since the definition of 2D orbital altermagnetism concerns magnetic states with a nontrivial local orbital-moment pattern, we do \emph{not} consider the nonmagnetic realization (type-II MSG) of gray point groups.
Therefore, throughout the screening below, whenever a gray (type-II) magnetic point group is encountered, it is implicitly understood as the realization compatible with a type-IV MSG (i.e., with $\tilde{\mathcal{T}}=\{1|\boldsymbol{\tau}\}\mathcal{T}$ rather than $\{1|\mathbf{0}\}\mathcal{T}$).

\paragraph*{(ii) Only operations preserving the $k_z=0$ plane.}
Since we target a two-dimensional orbital texture characterized on the $k_z=0$ plane, we only keep symmetry operations that leave this plane invariant. Concretely, we consider the subgroup of operations $g$ such that
\begin{equation}
g:\ (k_x,k_y,0)\mapsto (k_x',k_y',0).
\end{equation}
Operations that mix $k_z$ with in-plane components are discarded for the purpose of the 2D classification.

\paragraph*{(iii) Existence of a symmetry that flips the out-of-plane orbital moment.}
To realize the defining compensated pattern (zero net out-of-plane moment but nonzero local $M_z$), the symmetry group must contain at least one operation that maps the two sublattices onto each other while reversing the sign of the out-of-plane axial moment. In other words, we require the presence of a symmetry element $g$ such that
\begin{equation}
M_z(\rr) \xrightarrow{g} -\, M_z(g\rr),
\end{equation}
which guarantees cancellation of the net $z$-component upon summation over the symmetry-related sublattices (or loop-current substructures), while still allowing $M_z(\rr)\not\equiv 0$.

\paragraph*{(iv) Exclusion of symmetries that force $L_z(\kk)$ to vanish.}
Finally, we exclude magnetic point groups containing any (antiunitary) operation that enforces the constraint
\begin{equation}
L_z(\kk) = -L_z(\kk),
\end{equation}
which would symmetry-force $L_z(\kk)\equiv 0$ on the $k_z=0$ plane and thus forbid a finite orbital texture. In practice, this removes cases where an antiunitary symmetry maps $\kk\to\kk$ while flipping the sign of the axial out-of-plane moment (for example, $\mathcal{P}\mathcal{T}$ or $C_{2z}\mathcal{T}$, as discussed above).

\paragraph*{(v) Conventions and coordinate choices.}
Throughout this work, symmetry operations are taken in their standard settings with conventional axis orientations.
For a given material, the crystallographic axes may be rotated relative to these standard conventions, and a corresponding change of coordinates may be required to apply the screening rules to a specific structure.
Such material-dependent axis redefinitions do not change the symmetry logic, but they can affect which component is identified as the out-of-plane direction in the chosen coordinate system.
A case-by-case coordinate alignment is therefore left for future, system-specific analyses and is not pursued in detail here. For example, consider the magnetic point group $m'm2'$. In the standard convention one may represent its generators as $m_x\mathcal{T}$, $m_y$, and $C_{2z}\mathcal{T}$; in this orientation $C_{2z}\mathcal{T}$ can map $\kk\!\to\!\kk$ while reversing the axial out-of-plane moment, thereby symmetry-forcing $L_z(\kk)$ to vanish and excluding 2D orbital altermagnetism.
However, under a different (rotated) axis assignment for the same abstract group, one may instead take $m_z\mathcal{T}$, $m_x$, and $C_{2y}\mathcal{T}$ as generators; in that coordinate choice, the corresponding constraints no longer forbid a finite $L_z(\kk)$ texture, and 2D orbital altermagnetism becomes symmetry-allowed.

Applying the above criteria yields a restricted set of magnetic point groups compatible with two-dimensional orbital altermagnetism, which we summarize in Table~\ref{tab:mpg_2d_oam} and discuss below. Table~\ref{tab:mpg_2d_oam} is the same as Table I in the main text.

\begin{table}[h!]
\centering
\caption{Magnetic point groups (MPGs) that allow 2D orbital altermagnetism, together with the symmetry-enforced leading splitting wave of the orbital texture. The classification is given in the standard setting with conventional axis orientations; for a specific material, an appropriate rotation of the coordinate axes may be required.}
\label{tab:mpg_2d_oam}
\begin{tabular}{c c c}
\hline\hline
Leading wave & Allowed magnetic point groups (MPGs)  \\ \\
\hline
\textit{p}-wave & $\mathrm{1.1',\ 2.1,\ 2.1',\ m.1,\ m.1'}$ \\
\textit{d}-wave & $\mathrm{2/m.1,\ 222.1,\ mm2.1,}$\newline
         $\mathrm{mmm.1,\ 4',\ -4',\ 4'/m,\ 4'22',\ 4'm'm,\ -4'2'm,\ -4'2m',\ 4'/mm'm}$ \\
\textit{g}-wave & $\mathrm{422.1,\ 4mm.1,\ -42m.1,\ 4/mmm}$ \\
\textit{f}-wave & $\mathrm{3.1',\ 32.1,\ 32.1',\ 3m.1,\ 3m.1',\ -6.1',\ -6',\ -6m2.1,\ -6m2.1',\ -6'm'2,\ -6'm2'}$ \\
\textit{i}-wave & $\mathrm{-3m.1,\ 622.1,\ 6mm.1,\ 6/mmm.1}$ \\
\hline\hline
\end{tabular}
\end{table}

\subsection{Magnetic point groups allowing the 2D orbital-altermagnetic ferromagnet}

To identify the magnetic point groups that permit 2D orbital altermagnetic ferromagnets, we impose the following criteria: (i) the group must be compatible with ferromagnetism; (ii) it must contain at least one symmetry operation that reverses the $z$-component of magnetization (see Fig.~\ref{fig:example}); (iii) both $\mathcal{PT}$ and $C_{2z}\mathcal{T}$ symmetries must be excluded.
Applying these screening criteria yields only six compatible magnetic point groups: $2$, $m$, $m_z^\prime$, $2/m$, $m_z^\prime m 2^\prime$, and $m^\prime m_z^\prime 2$. Among these, only the $2/m$ group exhibits even-parity altermagnetism, which is invariant under $\kk \to -\kk$ (see Table~\ref{tab:MPG}).


\begin{figure}
\centering
\includegraphics[width=0.5\linewidth]{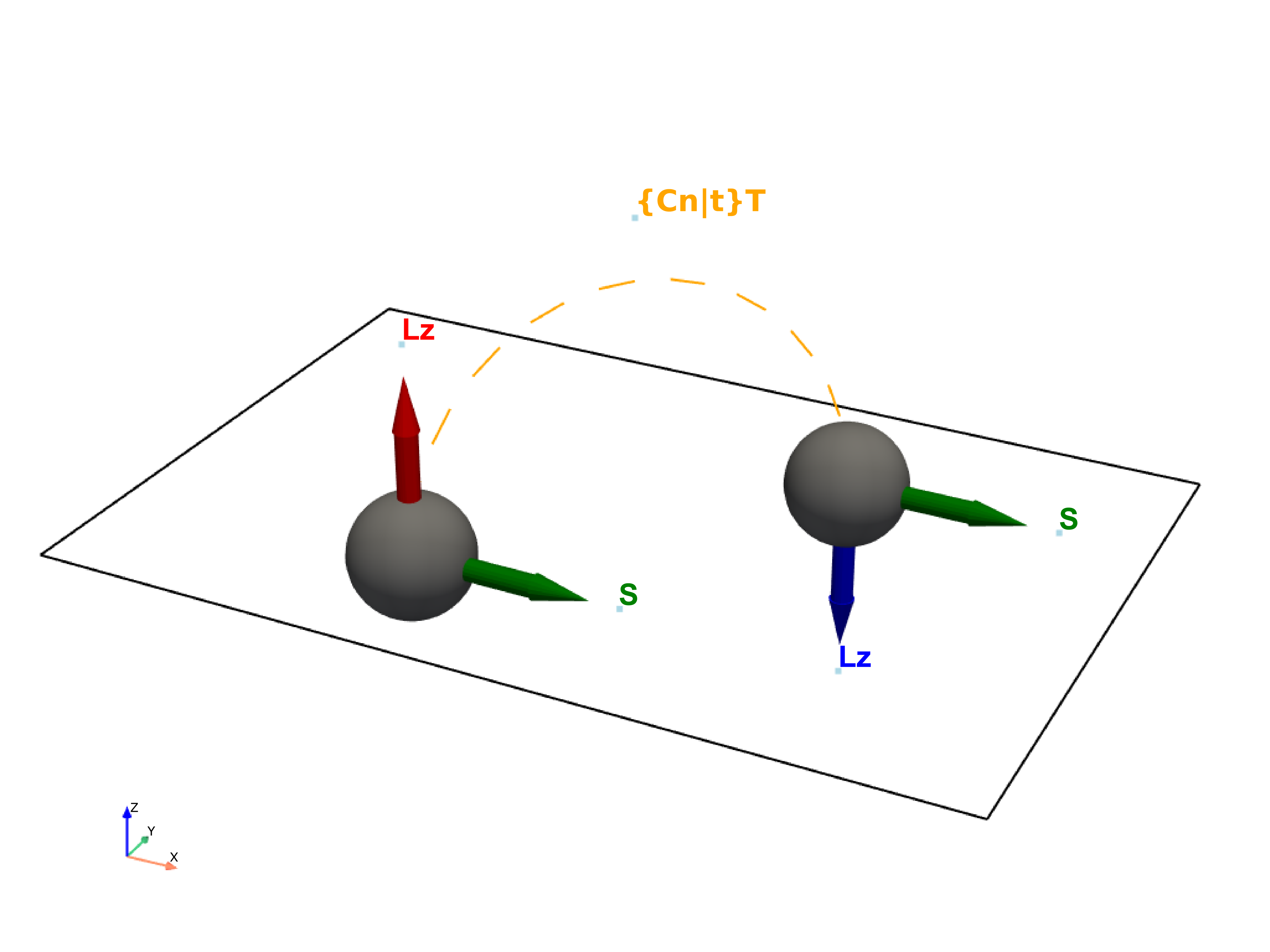}
\caption{\label{fig:example} A schematic diagram illustrates how an in-plane ferromagnet can be used to achieve an out-of-plane orbital antiferromagnet. A combined $\{C_n|t\}\mathcal{T}$ symmetry links to sites with opposite orbital angular momentum while maintaining the in-plane spin. Here, the arrangement of other non-magnetic elements is omitted, though they are typically required for the breaking of the $\mathcal{PT}$ or $\tau\mathcal{T}$ symmetry.}
\end{figure}

\begin{table}[h!]
\centering
\caption{Magnetic point groups allowing 2D orbital altermagneic ferromagnets. The criteria are: (i) compatibility with ferromagnetism, (ii) presence of at least one operation reversing the $z$-component of magnetization, and (iii) exclusion of both $\mathcal{PT}$ and $C_{2z}\mathcal{T}$ symmetries. The parity column indicates the parity under inversion $\kk\to -\kk$.}
\label{tab:MPG}
\begin{tabular}{c c c}
\hline\hline
Magnetic point group & Representative operations  & Parity \\ \\
\hline
$2$ & $C_{2y}$ & Odd \\
$m$ & $m_x$ &  Odd \\
$m_z^\prime$ & $m_z\mathcal{T}$ &  Odd \\
$2/m$ & $C_{2x}$, $m_x$ &  Even \\
$m_z^\prime m 2^\prime$ & $m_z\mathcal{T}$, $m_x$, $C_{2y}\mathcal{T}$ &  Odd \\
$m^\prime m_z^\prime 2$ & $m_x\mathcal{T}$, $m_z\mathcal{T}$, $C_{2y}$ &  Odd \\
\hline\hline
\end{tabular}
\end{table}

\subsection{Role of spin-orbit coupling in 2D orbital altermagnetism}

We emphasize that spin--orbit coupling (SOC) is central for the two-dimensional setting considered in this work. In the absence of SOC, spin and orbital sectors are decoupled. As a result, symmetry operations acting in spin space influence the orbital magnetic moment only through whether the associated symmetry is unitary or antiunitary when projected onto the orbital sector. In particular, for collinear or coplanar magnetic configurations, one can define an \emph{effective orbital time-reversal-like} antiunitary symmetry, which enforces the local orbital moment in real space to vanish identically. Therefore, such SOC-free collinear/coplanar states do not satisfy our definition of orbital altermagnetism, which requires $M_{\rm orb}(\rr)\not\equiv 0$.

More specifically, the real-space orbital magnetic moment $M_{\rm orb}(\rr)$ (an axial vector) transforms under a general spin space-group operation $\{W|R|\mathbf{t}\}$ as~\cite{cam5007}
\begin{equation}
\label{eq:Morbr_transform}
M_{\rm orb}(\rr)=\det(W)\,\det(R)\,R\cdot M_{\rm orb}\!\left(\{R|\mathbf{t}\}^{-1}\rr\right),
\end{equation}
where $\det (W)=-1$ can be viewed as an effective orbital time-reversal symmetry, $R$ is a proper or improper rotation, and $\mathbf{t}$ is a translation. Equation~\eqref{eq:Morbr_transform} implies that if the symmetry group contains an antiunitary operation of the form $\{W|1|\mathbf{0}\}$ with $\det(W)=-1$ that leaves $\rr$ invariant, then
\begin{equation}
M_{\rm orb}(\rr)=-\,M_{\rm orb}(\rr)=0,
\end{equation}
so the local orbital moment is symmetry-forced to vanish everywhere. Such operations can be viewed as an ``orbital time-reversal'' symmetry in the SOC-free limit ~\cite{PhysRevB.109.094438}.

For collinear magnetic configurations with spins aligned along the $z$ axis, the effective orbital time-reversal symmetry can be written as $\{\mathcal{T}U_{xy}(\pi)|1|\mathbf{0}\}$, consisting of time reversal followed by a $\pi$ spin rotation about an arbitrary in-plane axis. For coplanar structures with spins lying in the $xy$ plane, the corresponding effective orbital time-reversal symmetry is $\{\mathcal{T}U_{z}(\pi)|1|\mathbf{0}\}$, which transforms a spin moment $(S_x,S_y,S_z)$ to $(S_x,S_y,-S_z)$~\cite{PhysRevX.14.031037}. Consequently, \emph{without SOC, orbital altermagnetism is only possible for noncoplanar magnetic structures}, where no such effective orbital time-reversal symmetry exists. Under this condition, the main role of SOC is to lift the effective orbital time-reversal-type symmetry that would otherwise forbid orbital altermagnetism in collinear or coplanar spin structures, thereby allowing a nonzero orbital texture while leaving the final symmetry classification to be determined by the magnetic point group of the SOC-included system.

Finally, we note an additional simplification in the SOC-free limit.
Because spin-space operations $W$ affect the orbital magnetic moment only through whether $\det(W)=+1$ or $-1$, their action on $M_{\rm orb}$ reduces to a binary classification rather than depending on the detailed spin rotation itself.
As a consequence, any nontrivial spin point group can be associated with a corresponding magnetic point group that captures the induced constraints on the orbital sector (see Table 7 in ~\cite{10.21468/SciPostPhys.18.3.109}).
Therefore, when SOC is neglected, one may determine whether an orbital altermagnetism is symmetry-allowed by identifying the magnetic point group corresponding to the noncoplanar configuration's nontrivial spin point group and then applying our magnetic-point-group screening results.

This conclusion, however, holds under the assumption that the orbital sector itself does not introduce any additional time-reversal-symmetry breaking. In other words, it applies when time-reversal symmetry is broken only by the spin sector, so that the relevant symmetry analysis can be based entirely on the spin-only group and its associated nontrivial spin point group. A more subtle situation arises when the orbital sector itself also breaks time-reversal symmetry, for example, through loop-current-like order or other non-spin mechanisms. In that case, some effective orbital time-reversal symmetries that would otherwise be present, especially in collinear or coplanar spin structures, may already be removed even without SOC. One must then first determine whether the effective orbital time-reversal symmetry remains preserved after including the orbital-sector time-reversal symmetry breaking. If it is broken, one should identify the surviving subgroup of the original symmetry, determine the corresponding reduced nontrivial spin point group, and then map it to the associated magnetic point group. Once this reduced symmetry is established, the same magnetic-point-group screening criterion can again be applied to determine the final symmetry-allowed orbital texture. In this sense, SOC is not the only possible route to lift the forbidding symmetry constraint, but it is the most common and physically natural one in realistic materials, whereas other mechanisms are usually limited to more special model settings or material platforms.

This viewpoint can be summarized more systematically using the concept of the oriented spin space group (OSSG)~\cite{liu_symmetry_2026}, namely an SSG with a fixed magnetic orientation. In the SOC-free case, one should first determine the relevant OSSG, including any additional time-reversal-symmetry breaking from the orbital sector if present. For orbital-sector screening, the spin-space part enters only through whether \(\det(W)=+1\) or \(-1\), so the OSSG can be reduced to an effective magnetic point group and the same magnetic-point-group screening criterion can then be applied. When SOC is included, the spin and lattice rotational parts are locked, and the OSSG reduces directly to the ordinary magnetic space group. In this language, the cases where SOC is genuinely essential are those in which the effective magnetic point group associated with the OSSG forbids orbital altermagnetism, while the SOC-reduced magnetic symmetry allows it.

\begin{figure}[h]
\centering
\includegraphics[width=0.8\linewidth]{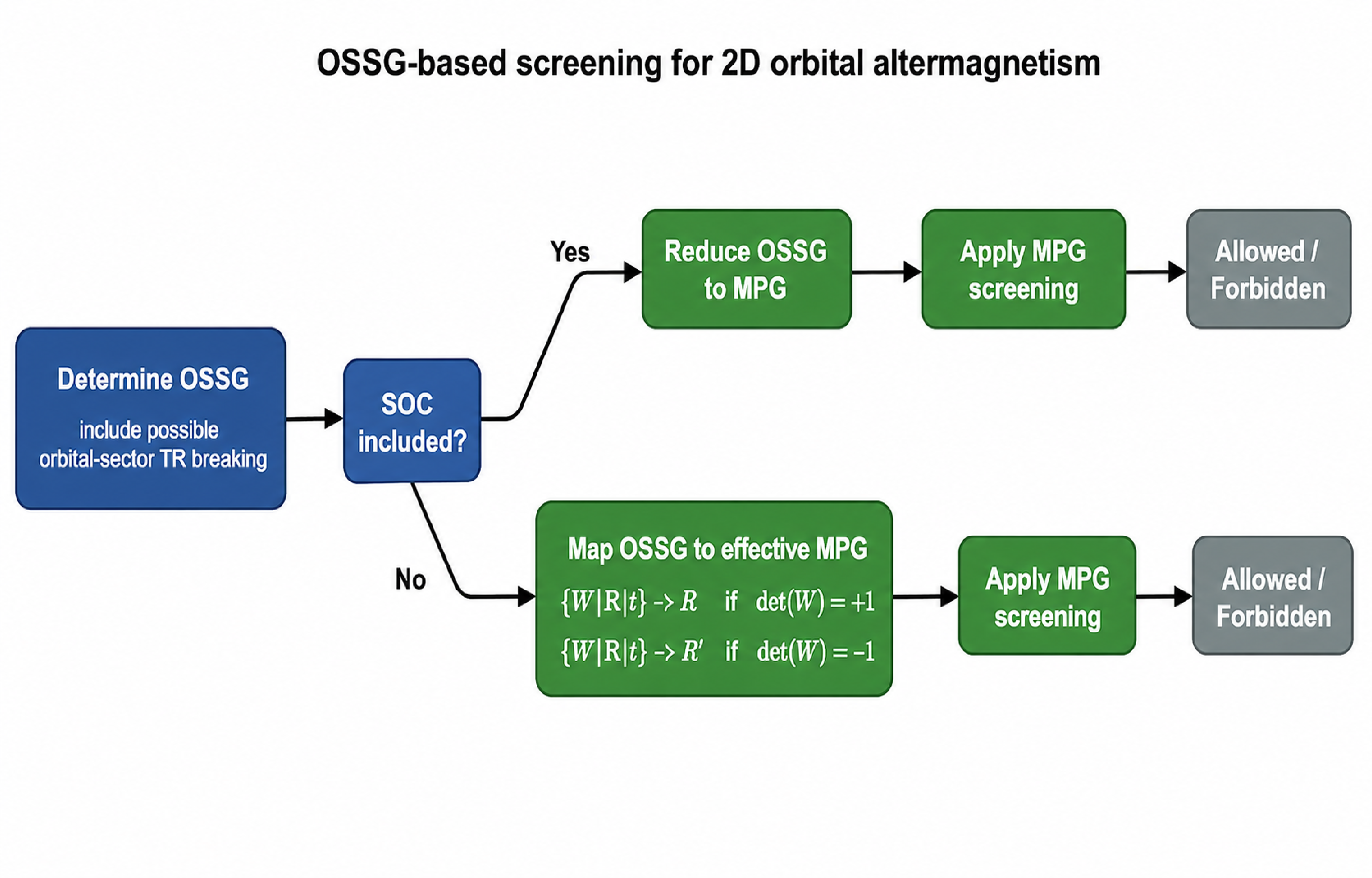}
\caption{\label{flow} Schematic illustration of the OSSG-based screening procedure for two-dimensional orbital altermagnetism. With SOC, the oriented spin space group (OSSG) reduces directly to the corresponding magnetic point group (MPG). Without SOC, the OSSG is first mapped to an effective MPG through the sign of \(\det(W)\), which determines whether the spin-space operation preserves or reverses the orbital moment. The resulting MPG is then used for the final symmetry screening.}
\end{figure}

\section{First-Principles Calculation Method}
First-principles calculations are performed within the framework of density-functional theory (DFT) using the projector augmented-wave method and the Perdew-Burke-Ernzerhof-type exchange-correlation functional~\cite{PhysRevLett.77.3865}, as implemented in the Vienna \textit{ab initio} simulation package (VASP)~\cite{KRESSE199615}. We set the cutoff energy of the plane-wave basis to be 520 eV and adapted a 15$\times$15$\times$1 $\Gamma$-center k-mesh for the BZ sampling. SOC is included self-consistently in the calculations. To eliminate the spurious interaction between the 2D material and its images, a vacuum region of more than 12~\AA~along the $z$ direction is set in the calculation. All crystal structures are fully optimized until the forces are less than $10^{-3}$ eV/\AA~and the change of the total energy is less than 0.01 meV. The correlation effect for the Cr 3$d$ electrons and Mo 4$d$ electrons is treated by the DFT+$U$ method with $U=3.55$ and $2.4$ eV~\cite{10.1063/5.0147450, CALDERON2015233}, respectively. To study the in-plane ferromagnetic configuration, we further performed constrained-moment calculations in VASP using the tags \texttt{I\_CONSTRAINED\_M} and \texttt{M\_CONSTR} to align the spin moments along a chosen in-plane direction. To study the band geometric quantities such as the Berry curvature and orbital magnetization, we construct maximally localized Wannier functions (MLWFs) and derive a Wannier-based \textit{ab initio} tight-binding model for each structure using the Wannier90 code~\cite{MOSTOFI2008685}. Based on the Wannier representation,  we calculate the Berry curvature $\Omega(\kk)$ and orbital magnetization of all states. All input structures, calculation parameters, and Wannier90 projectors used in this work are openly available in the Zenodo repository at \href{https://doi.org/10.5281/zenodo.19602057}{10.5281/zenodo.19602057}.

\section{Numerical results of other candidate materials for orbital altermagnetism}

\begin{figure}[h]
\centering
\includegraphics[width=0.8\linewidth]{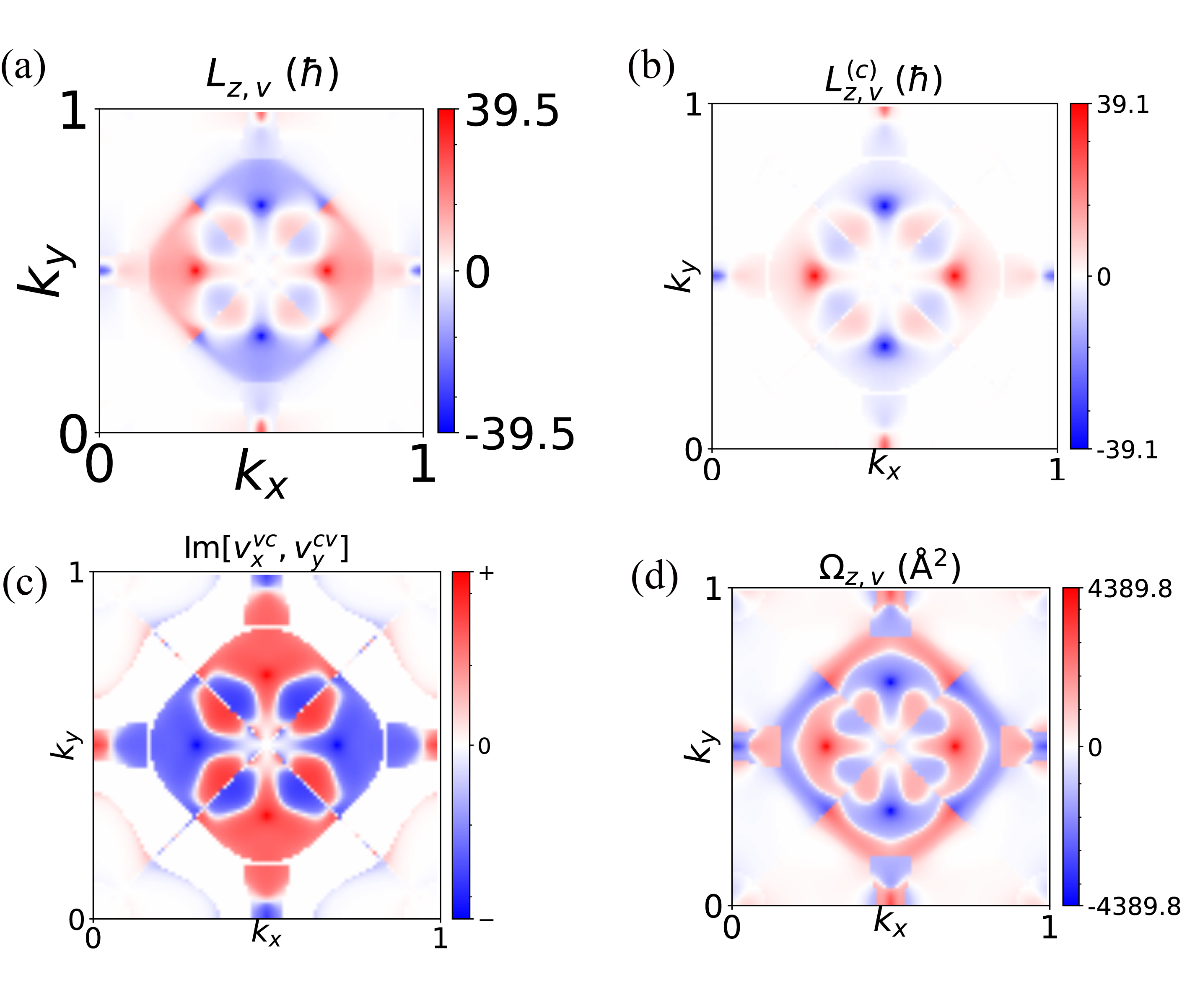}
\caption{\label{figr_quantites} Momentum-space distributions of quantities related to the orbital angular moment for the highest valence band ($n=v$) and the lowest conduction band ($m=c$) of the MoO monolayer. (a) Distribution of the total $L_{z,v}$ in the Brillouin zone. (b) Distribution of the interband-resolved contribution $L_{z,v}^{(c)}$. (c) Distribution of $\textrm{Im}[v_x^{vc},v_y^{cv}]$, showing the nodal lines separating regions of opposite sign. (d) Distribution of the Berry curvature $\Omega_{z,v}$ for comparison.}
\end{figure}

\begin{figure}[h]
\centering
\includegraphics[width=0.8\linewidth]{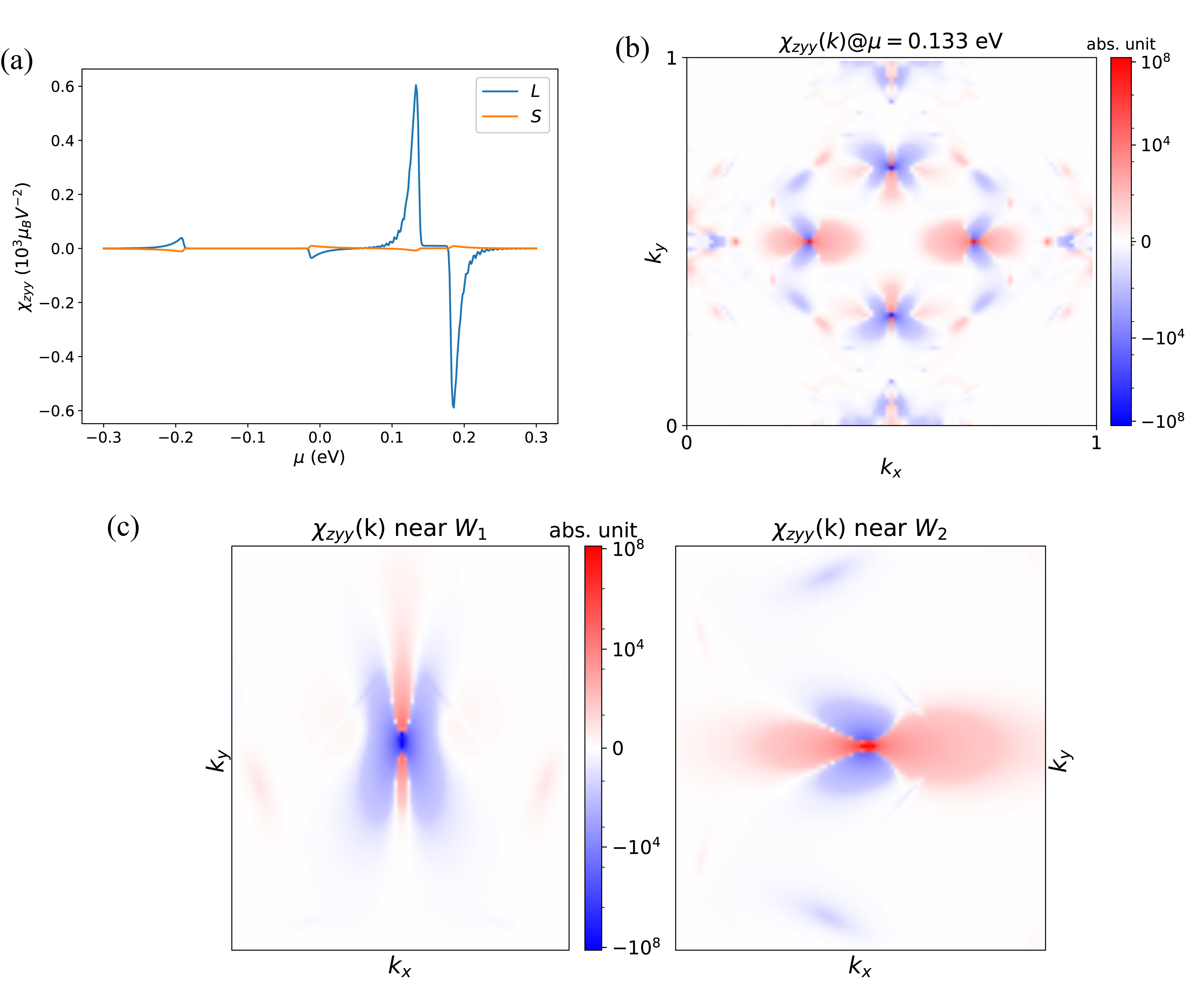}
\caption{\label{fig:w1w2} (a) $\chi_{zyy}$ versus the chemical potential $\mu$ for monolayer MoO based on the first-principle calculations. (b) The distribution of $\chi_{zyy}(\kk)$ across the Brillouin zone, with the $k_x$ and $k_y$ expressed in reduced coordinates. (c) The distribution of $\chi_{zyy}(\kk)$ near the $W_1$ and $W_2$, where $\mu$ is fixed at 0.133 eV.}
\end{figure}

\subsection{Current-induced orbital magnetization in MoO monolayer}

We first present the $\bm{\kk}$-space distribution of quantities related to the orbital angular moment. Figure~\ref{figr_quantites}(b) shows the $L_z$ of the highest valence band $v$ mainly originates from the lowest conduction band $c$. Here we define:
\begin{equation}
\begin{split}
&M_{n\kk}^{(m)} = -\frac{e}{2\hbar c}\,\mathrm{Im}
\frac{\langle u_{n\kk}|\hat{\mathbf{v}}|u_{m\kk}\rangle \times
\langle u_{m\kk}|\hat{\mathbf{v}}|u_{n\kk}\rangle}
{\varepsilon_{n\kk}-\varepsilon_{m\kk}} \qquad (m\neq n),\\
&[M_{n\kk}^{(m)}]_z =-g_L\mu_BL_{z,n}^{(m)}(\kk).
\end{split}
\end{equation}
To investigate the origin of the frequent sign changes of $L_z$ in k-space, we focus on the numerator in the expression Im[$v_{x}^{vc}$,$v_{x}^{vc}$]=Im($v_{x}^{vc}$$v_{y}^{cv}$-$v_{y}^{vc}$$v_{x}^{cv}$), and plot its $\kk$-space distribution for the $c$ and $v$ bands, as shown in Fig.~\ref{figr_quantites}(c). In momentum space, Im[$v_{x}^{vc}$,$v_{x}^{vc}$] typically changes sign across the lines or regions where it vanishes. These zero-value contours divide the Brillouin zone into several sectors, and crossing such a line naturally leads to a reversal of the sign of the orbital magnetic moment. Similarly, the numerator of the Berry curvature has a comparable form. Therefore, we also plotted the Berry curvature distribution in $\kk$-space, which shows a similar sign-reversal behavior.

Spin–orbit coupling opens a gap at the Weyl points $W_1$ and $W_2$~\cite{PhysRevB.107.214419}. However, in the vicinity of these nearly degenerate points, the orbital angular momentum $L_z$ remains strongly enhanced. Under an external field, these points contribute much more significantly to orbital magnetization than spin polarization. We calculated the nonlinear current-induced spin polarization (CISP) and current-induced orbital polarization (CIOP) in the MoO monolayer, as illustrated in Fig.~\ref{fig:w1w2}(a). The results reveal that the orbital polarization is substantially stronger than the spin polarization. As shown in Fig.~\ref{fig:w1w2}(b), $W_1$ and $W_2$ points represent the hotspot regions in the $\kk$-space distribution of $\chi_{zyy}(\kk)$. While the overall distribution follows the symmetry of the system, the local behaviors around $W_1$ and $W_2$ are quite different, as shown in Fig.~\ref{fig:w1w2}. In particular, although the $L_z$ distribution is linked by $C_{4z}\mathcal{T}$, the velocity matrix elements $v_y$ at $W_1$ and $W_2$ lack such a symmetry connection, preventing cancellation between their contributions.

Figures~\ref{fig:w1w2}(b,c) further demonstrate that the intrinsic CIOP is highly sensitive to band-energy differences, with the dominant contribution arising from the Fermi surface. This indicates that metallic or semimetallic 2D altermagnets are especially promising candidates for realizing large CIOP. Symmetry analysis shows that in MoO the out-of-plane CIOP exhibits a $\pi$-periodic angular dependence: $\chi_{zxx}=-\chi_{zyy}$ and $\delta X_z=\chi_{zxx}\cos2\theta$, where the applied electric field is parametrized as $(E_x,E_y)=E(\cos\theta,\sin\theta)$.

Our calculations also indicate that applying 5\% tensile strain in monolayer MoO shifts the $W_1$ and $W_2$ points close to the Fermi energy~\cite{PhysRevB.107.214419}. Since the SOC-induced gap is small, the associated orbital polarization $L_z$ is expected to remain robust. This suggests that pressure or strain engineering offers a practical experimental route to probing strong orbital polarization effects in the MoO monolayer.

\subsection{Monolayer altermagnet CrO}
Figure~\ref{fig:cro} presents the crystal structure, electronic band structure, and $L_z$ distribution in momentum space for monolayer CrO, demonstrating its characteristic orbital altermagnetic features.

\begin{figure*}[h]
\centering
\includegraphics[width=1.0\linewidth]{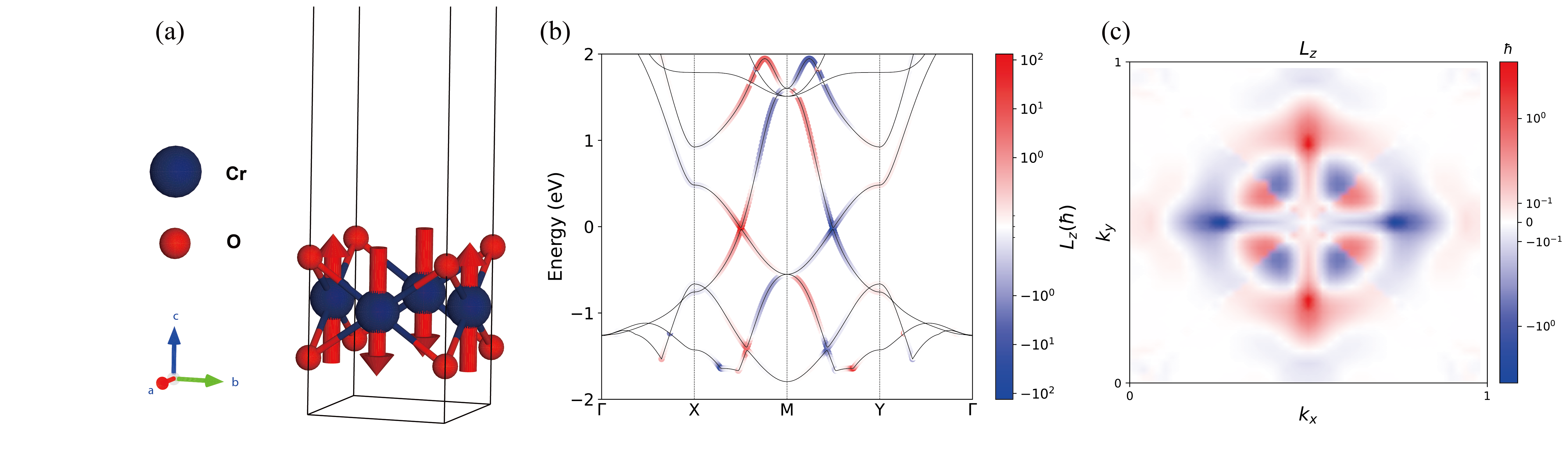}
\caption{\label{fig:cro} (a) Side view of the lattice structure (red O atoms, blue Cr atoms). (b) Band structures of CrO, where the color represents the magnitude of $L_z$ for the corresponding band and $k$-point. (c) The distribution of $L_{z}(\kk)$ across the Brillouin zone, with the $k_x$ and $k_y$ expressed in reduced coordinates. To better visualize the variation of $L_z$, all figures related to the $L_z(\kk)$ distribution use a logarithmic color scale.
}
\end{figure*}

\subsection{Monolayer ferromagnet VS$_2$}
The magnetic space group of monolayer ferromagnet VS$_2$ is $2/m$. As illustrated in Fig.~\ref{fig:vs}, the inter-site currents between V and S atoms give rise to closed current loops within rhombic plaquettes consisting of two V atoms and their two adjacent S atoms. These current loops alternate in direction, with counter-circulating loops related by either the $C_{2x}$ rotation or the $m_x$ mirror symmetry.

\begin{figure*}
\centering
\includegraphics[width=1.0\linewidth]{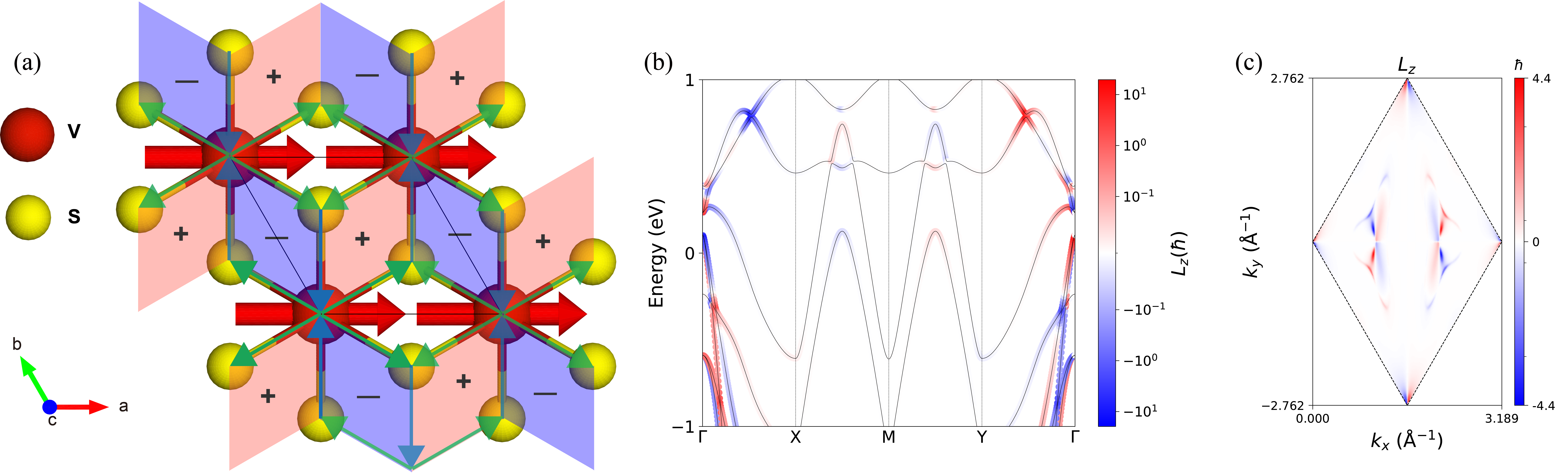}
\caption{\label{fig:vs} (a) Crystal structure of monolayer VS$_2$ from the top view. The spin magnetic moments localized on the V atoms are annotated with arrows. Some inter-site currents between V and S atoms are marked with colored arrows. Blue corresponds to a current magnitude of 0.44 $\mu$A, and green corresponds to 0.22 $\mu$A. (b) Band structures of VS$_2$, where the color represents the magnitude of $L_z$ for the corresponding band and $k$-point. (c) The distribution of $L_{z}(\kk)$ for the band 14 across the Brillouin zone. The boundaries of the Brillouin zone are marked with black dashed lines. To better visualize the variation of $L_z$, all figures related to the $L_z(\kk)$ distribution use a logarithmic color scale.
}
\end{figure*}

\subsection{Monolayer ferromagnet CuBr$_2$ with s$\parallel$x}
Figure~\ref{fig:cubrx} shows the calculated band structure of CuBr$_2$ together with the corresponding $L_z$ distribution in momentum space for the configuration with spin polarization along the $x$-axis. Figure~\ref{fig:ciopcubr} shows the curves of CIOP (CISP) in CuBr$_2$ for different spin orientations.

\begin{figure*}
\centering
\includegraphics[width=1.0\linewidth]{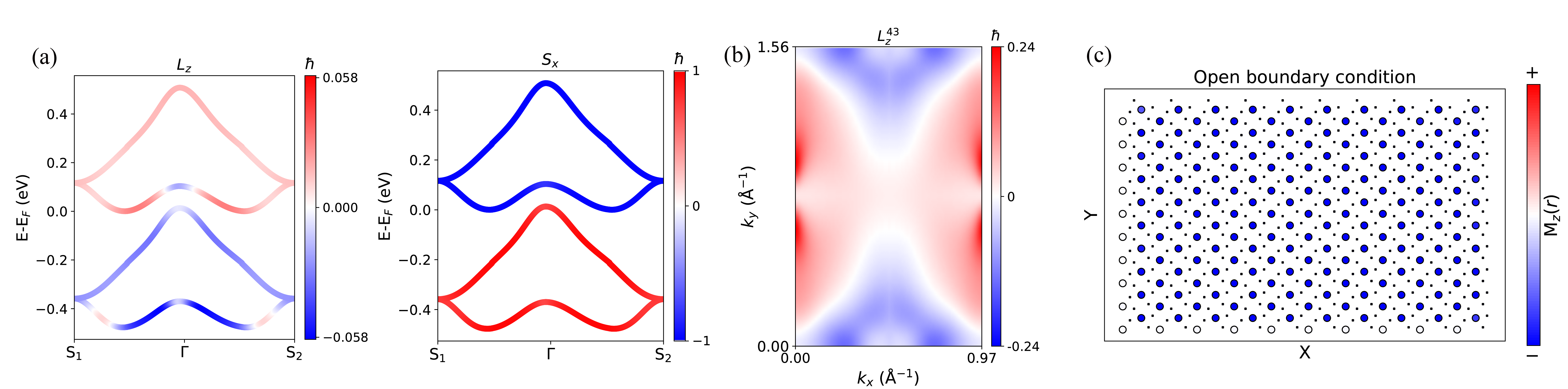}
\caption{\label{fig:cubrx} (a) Band structures of CuBr$_2$ when spin is along x-axis, where the color represents the magnitude of $L_z$ and $S_x$ for the corresponding band and $k$-point.Here S$_1$=(0.5,0.5) and S$_2$=(0.5,-0.5). (b) The distribution of $L_z(\kk)$ for the band 43 across the Brillouin zone. (c) The real-space distribution of orbital magnetic moments in a 10$\times$10 system under open boundary conditions. Here we choose $\mu$=0 eV.}
\end{figure*}

\begin{figure*}
\centering
\includegraphics[width=1.0\linewidth]{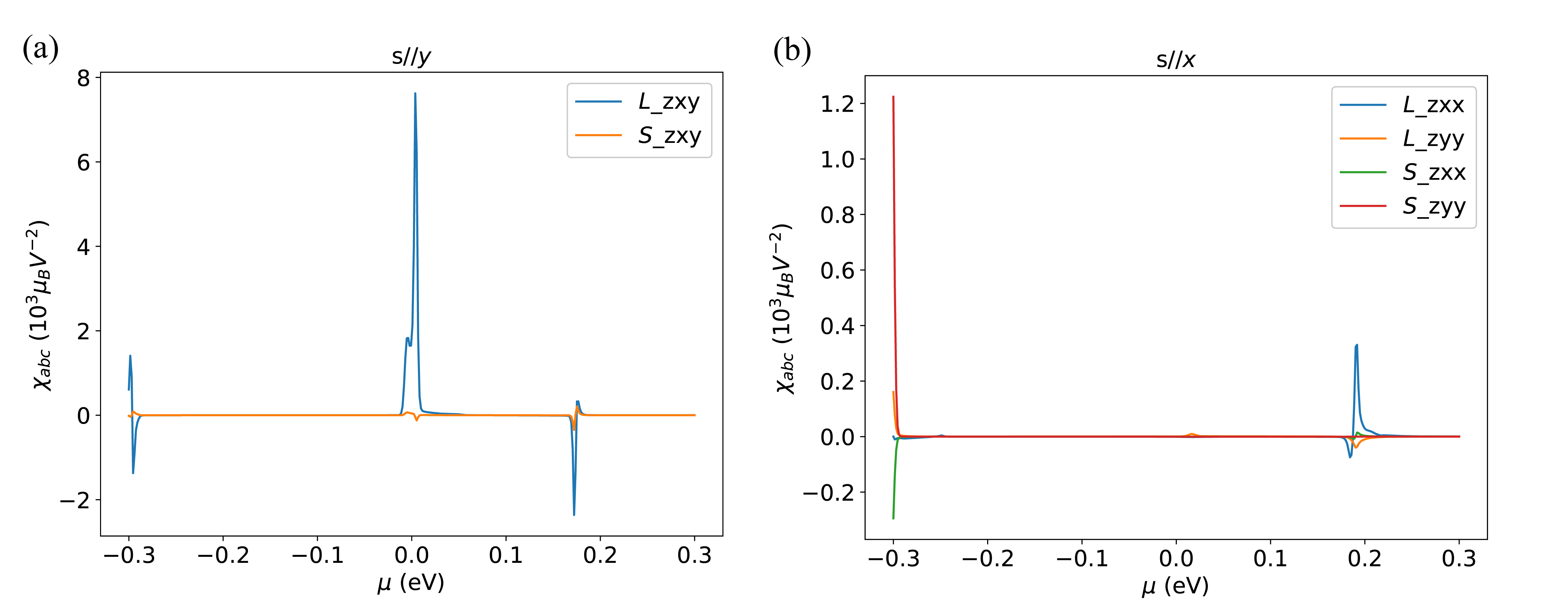}
\caption{\label{fig:ciopcubr} $\chi_{abc}$ versus the chemical potential $\mu$ for monolayer CuBr$_2$ based on the first-principle calculations. (a) corresponds to spin polarization along $y$-axis, while (b) shows the case for $x$-axis polarization.}
\end{figure*}

\section{Discussion on experimental detection of orbital altermagnetism}
In this section, we briefly discuss the possible experimental detection of the orbital altermagnetism, based on the recent developments on
circular dichroic (angle-resolved) photoemission and nanoscale nitrogen vacancy (NV) centre quantum sensors.

Firstly, using circular dichroism (CD) in soft x-ray angle-resolved photoemission spectroscopy (ARPES), Brinkman, \textit{et al.} reports the observation of the bulk orbital-angular-momentum texture and monopolelike orbital-momentum locking in the bulk electronic structure of chiral crystal CoSi \cite{PhysRevLett.132.196402}. By introducing the intrinsic chiral circular dichroism as a differential photoemission observable and a natural probe of chiral electron states, this approach can be effectively applied to detect chiral loop current-induced orbital texture in $\kk$-space. Accordingly, we expect that the d-wave altermagnetic order of orbital-angular-momentum texture can be measured using circular dichroism in bulk-sensitive soft x-ray ARPES, to provide additional evidence for our proposed orbital altermagnetic materials.

Secondly, for materials where orbital magnetization arises from loop currents or intra-unit-cell circulation, nanoscale magnetic field sensors such as nitrogen-vacancy (NV) centers in diamond offer a promising complementary technique. More recently, Xie and Nagaosa \cite{NVcenter2025Nagaosa} demonstrated that NV magnetometry can detect the ``local'' stray magnetic fields generated by the fluctuation of orbital loop currents. This approach is particularly suitable for systems with relatively large sublattice separations, where the alternating orbital moments would generate measurable spatial variations in the stray magnetic field. In our context, scanning NV imaging could thus be employed to map the local magnetic field texture above the surface and indirectly confirm the predicted orbital altermagnetic pattern.

\end{widetext}


\end{document}